\documentclass[aps,showpacs,superscriptaddress,groupedaddress]{revtex4-1} 

\pdfoutput=1

\usepackage{amsmath}
\usepackage{amssymb}

\usepackage{graphicx}
\usepackage{subfigure}

\usepackage{dcolumn}
\usepackage{bm}
\usepackage{array}
\usepackage{color}

\graphicspath{{../figures/}}

\begin{document}

\title{
Generalized $k$-core pruning process on directed networks
}
\author{
Jin-Hua Zhao
\footnote{E-mail: \textbf{zhaojh190@gmail.com}}
}
\affiliation{
Department of Applied Science and Technology,
Politecnico di Torino,\\
Corso Duca degli Abruzzi 24,
10129 Torino,
Italy
}
\date{June 24, 2017}


\begin{abstract}
The resilience of a complex interconnected system
concerns the size of the macroscopic functioning node clusters
after external perturbations based on a random or designed scheme.
For a representation of the interconnected systems
with directional or asymmetrical interactions among constituents,
the directed network is a convenient choice.
Yet how the interaction directions affect the network resilience
still lacks thorough exploration.
Here, we study the resilience of directed networks
with a generalized $k$-core pruning process as a simple failure procedure
based on both the in- and out-degrees of nodes,
in which any node with an in-degree $< k_{in}$ or an out-degree $< k_{ou}$
is removed iteratively.
With an explicitly analytical framework,
we can predict the relative sizes of residual node clusters
on uncorrelated directed random graphs.
We show that
the discontinuous transitions rise for cases with $k_{in} \geq 2$ or $k_{ou} \geq 2$,
and the unidirectional interactions among nodes
drive the networks more vulnerable
against perturbations based on in- and out-degrees separately.
\end{abstract}

\maketitle

\tableofcontents

\clearpage

\section{Introduction}


Resilience of the real interconnected systems
(such as infrastructural, ecological, and financial systems)
against external perturbations
is among the fundamental themes in the understanding of complex systems,
especially the phenomena of sharp transitions or tipping points
in their structural and also dynamical regimes
\cite{
Scheffer.etal-Science-2012}.
When it's adopted as a representation of nodes as constituents and edges as interactions among them,
the complex network theory
\cite{
Albert.Barabasi-RMP-2002,
Newman-SIAM-2003,
Boccaleti.Latora.Moreno.Chavez.Hwang-PhysRep-2006,
Dorogovtsev.Goltsev-RMP-2008}
in many cases offers
an analytical framework
for the resilience problem
as a percolation transition
\cite{
Stauffer.Aharony-1994}
which usually involves 
an emergence of macroscopic residual subgraph structure
against a removal procedure of nodes and edges.
Typical methods to explore the resilience of networks with undirected interactions are
the shrinkage of the giant connected component against a random node removal
\cite{
Albert.Jeong.Barabasi-Nature-2000,
Cohen.etal-PRL-2000,
Callaway.etal-PRL-2000},
the $K$-core percolation based on a degree-constrained pruning process
\cite{
Chalupa.Leath.Reich-JPhysC-1979,
Pittel.Spencer.Wormald-JCombTheor-1996,
Dorogovtsev.Goltsev.Mendes-PRL-2006,
Baxter.etal-PRX-2015},
the $K$-core percolation with an inducing effect on the intact (not failed) nodes
exerted by the collapsed ones
\cite{
Zhao.Zhou.Liu-NatCommun-2013},
the core percolation with a greedy leaf-removal procedure
\cite{
Liu.Csoka.Zhou.Posfai-PRL-2012},
and a node removal process based on articulation points
each of which can disintegrate a network into multiple components after its removal
\cite{
Liu.etal-NatCommun-2017}.
When we consider the networks with directional interactions,
we have examples as
the emergence of giant strongly connected components against random node removals
\cite{
Tarjan-SIAMJourComp-1972,
Dorogovtsev.Mendes.Samukhin-PRE-2001,
Schwartz.Cohen.benAvarham.Barabasi.Havlin-PRE-2002},
the core percolation
in specific contexts of defining leaves and node removals
\cite{
Liu.Csoka.Zhou.Posfai-PRL-2012,
AzimiTafreshi.etal-PRE-2013}.
With the notion of multilayer networks
\cite{
Boccaletti.etal-PhysRep-2014,
Kivela.etal-JCompNet-2014},
the effect of the coupling between nodes or node copies on the network resilience
are studied through the stability of connected components under a random node removal
\cite{
Buldyrev.etal-Nature-2010} and
a $K$-core-like pruning process
\cite{
AzimiTafreshi.etal-PRE-2014}.
The interaction direction
is further incorporated into the multilayer networks to explore their resilience
based on the size of the strongly connected components
against a random node removal
\cite{
Liu.Stanley.Gao-PNAS-2016}.
Besides the above research line of the network resilience in various contexts,
there is another one on the implication of interaction directions on networks
from the perspectives of
the structural organization
\cite{
Bianconi.Gulbahce.Motter-PRL-2008,
Goltsev.etal-PRL-2017}
and the dynamical processes on them
 \cite{
Son.etal-PRL-2009}.
Yet the intersection of the two research lines,
the effect of interaction directions on the network resilience,
is still far from being fully discussed from an analytical perspective
except for the above few cases,
on which we lay our focus in this paper.

Here we study the resilience of directed networks
through a node failure model
which can be considered as a generalized $k$-core pruning process.
In the failure scheme,
a node fails once it has too small an in-degree or an out-degree.
The motivation of the failure model is that,
in systems with directed or asymmetrical interactions such as those involving information flows
or purely with a principle of redundancy design,
the proper functioning of each constituent
can be assumed to be based on sufficient sizes
of both its neighbors
from which it receives interactions
and to which it delivers influences.
Formulating the failure model
into a pruning process on a directed network,
we randomly and iteratively remove any node
with an in-degree smaller than an  integer $k_{in} (\geq 0)$
or with an out-degree smaller than an integer $k_{ou} (\geq 0)$
along with all its adjacent arcs,
and we study the relative size of the macroscopic residual subgraph.
As we can see,
the above process is 
basically an extension of $K$-core pruning process on undirected networks
into the case with directed interactions.
This pruning process is initially discussed in
\cite{
Giatisdis.etal-ICDM-2011},
which lists the sizes of all the non-trivial residual structures
after the above pruning process on a real network
as a tool for data analysis.
Yet a detailed theoretical analysis of
the related percolation problem,
or the $(k_{in}, k_{ou})$-core percolation as we can simply put,
is still missing.
In this paper,
we consider the generalized $k$-core pruning process
and the related $(k_{in}, k_{ou})$-core percolation problem
as a solvable model for the resilience study of the directed networks,
and we mainly work on the derivation of an analytical framework
and the analysis of its transition behaviors.

There are two parts of our main results.
$(1)$
The $(k_{in}, k_{ou})$-core percolation problem
with $k_{in} \geq 2$ or $k_{ou} \geq 2$
shows abrupt transitions
on infinitely large directed random networks,
which can be proved with our analytical framework.
$(2)$
We compare the relative sizes of
the $(k_{in}, k_{ou})$-core on a directed random graph
with the $K$-core on its undirected counterpart
when the arc directions are totally ignored.
Based on the transition points of
two percolation problems,
we can see from an analytical perspective
that when a macroscopic $(k_{in}, k_{ou})$-core
is permitted in some case,
a $K$-core with a much larger $K$ than $k_{in}$ and $k_{ou}$
is possible,
indicating that the introduction of unidirectional interactions between nodes
can drive the connected systems more vulnerable
against external perturbations based on
in-degrees and/or out-degrees of nodes.

Here is the structure of the paper.
In Sec.\ref{sec:model},
we explain the pruning process and the percolation problem
in a general setup in networks
with both undirected and directed interactions.
In Sec.\ref{sec:theory},
we present an analytical theory for the problem
on random networks with only unidirectional interactions.
In Sec.\ref{sec:results},
we test the theory on model directed random networks and also real network data sets,
in which we also examine the discontinuity and the scaling property of the hybrid transitions.
In Sec.\ref{sec:conclusion},
we conclude the paper with a discussion.

\section{Model}
\label{sec:model}

\begin{figure}
\begin{center}
 \includegraphics[width = 0.70 \linewidth]{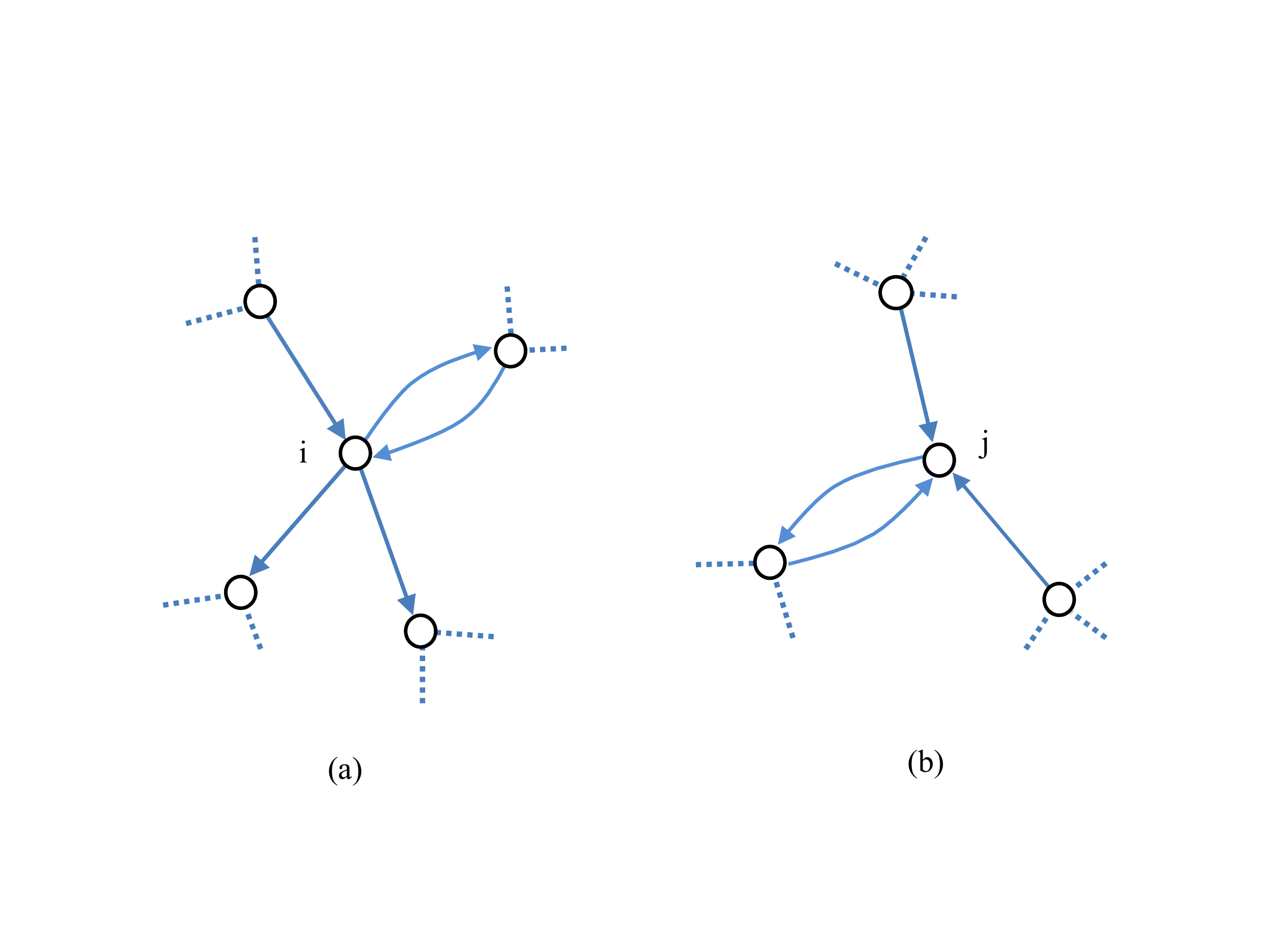}
\end{center}
\caption{
\label{fig:model}
Examples of the pruning process in the $(k_{in}, k_{ou})$-core percolation.
Here are two examples of elementary removal steps
in $(3, 2)$-core percolation.
The residual arcs are in solid lines,
and the dashed lines denote other connections.  
In (a), the node $i$ has $2$ in-coming arcs and $3$ out-going arcs.
In (b), the node $j$ has $3$ in-coming arcs and $1$ out-going arc.
Both the nodes $i$ and $j$ along with their adjacent arcs will be removed in the pruning process.}
\end{figure}
%

First we explain some notations for directed networks.
We consider a directed network instance $D = \{V, A\}$
with a node set V ($|V| = N$) and an arc set A ($|A| = M$),
correspondingly its arc density $c \equiv M / N$.
In the context of directed networks,
we adopt the arc density
(the arc-node ratio, or the number of connections divided by the number of nodes)
to describe the density of connections among nodes for a network.
We should mention that in the context of undirected networks,
a conventional notation of connection densities is the mean degree or the mean connectivity
(average number of connections adjacent to a node),
which is two times of the edge-node ratio
(also the number of connection divided by the number of nodes).
In order to avoid the confusion of notations,
we will specify the context of undirected networks
when we mention the mean degree or the mean connectivity.
A directional arc between two nodes in the network,
say $i$ and $j$,
is denoted as an ordered pair $(i, j)$
as an interaction or relation pointing from $i$ to $j$.
For an arc $(i, j)$,
$i$ is considered as the predecessor (an in-neighbor) of $j$,
and $j$ is considered as the successor (an out-neighbor) of $i$.
Correspondingly,
an arc $(i, j)$ is considered as 
an out-going arc for node $i$ and
an in-coming arc for node $j$.
For any node $i$,
all its in-neighbors constitute a set $\partial i^{+}$ with an in-degree $k_{+} (\equiv |\partial i^{+}|$), and
all its out-neighbors constitute a set $\partial i^{-}$ with an out-degree $k_{-} (\equiv |\partial i^{-}|$).
We define the degree distribution $P(k_{+}, k_{-})$ of a directed network $D$
as the probability that a randomly chosen node has $k_{+}$ in-neighbors and $k_{-}$ out-neighbors.
The arc density is thus
$c = \sum _{k_{+}, k_{-}} k_{+} P(k_{+}, k_{-}) = \sum _{k_{+}, k_{-}} k_{-} P(k_{+}, k_{-})$.
We further define two excess degree distributions.
For a randomly chosen arc $(i, j)$,
from node $i$ following the arc direction to node $j$,
the probability of node $j$ having $k_{+}$ in-neighbors and $k_{-}$ out-neighbors is $Q_{+}(k_{+}, k_{-})$;
from node $j$ following the opposite arc direction to node $i$,
the probability of node $i$ having $k_{+}$ in-neighbors and $k_{-}$ out-neighbors is $Q_{-}(k_{+}, k_{-})$.
We simply have
$Q_{+}(k_{+}, k_{-}) = k_{+} P(k_{+}, k_{-}) / c$ and
$Q_{-}(k_{+}, k_{-}) = k_{-} P(k_{+}, k_{-}) / c$.

On a directed graph $D = \{V, A\}$,
an initial fraction $p \in [0, 1]$ is defined in a starting step,
and a fraction $1 - p$ of nodes are randomly chosen
and further removed along with all their adjacent arcs.
An iterative pruning process is then carried out
as any node $i$ with an in-degree $k_{+}^{i} < k_{in}$ or an out-degree $k_{-}^{i} < k_{ou}$ is removed,
along with all its adjacent arcs.
The generalized $k$-core pruning process can be named as
the $(k_{in}, k_{ou})$-core pruning process.
We can see that if there are some nodes remained after the removal process,
any node $i$ in this subgraph has both an in-degree $k_{+}^{i} \geq k_{in}$
and an out-degree $k_{-}^{i} \geq k_{ou}$.
We simply call
the collection of the residual nodes and arcs
as the $(k_{in}, k_{ou})$-core or the core structure, and
the emergence of a macroscopic residual subgraph
as the $(k_{in}, k_{ou})$-core percolation.
See figure \ref{fig:model}
for some examples of the elementary pruning steps
in the $(k_{in}, k_{ou})$-core percolation.

The above model
is considered on networks with only directed arcs,
yet it can be also defined in a generalized case on networks with both undirected edges and directed arcs
\cite{
Boguna.Serrano-PRE-2005}.
In a mapping procedure,
each undirected edge is considered as a degenerate connection,
and is further split into two directed arcs in opposite directions.
For example,
an undirected edge as an unordered pair $\{i, j\}$ between nodes $i$ and $j$
is split into two directed arcs $(i, j)$ and $(j, i)$.
After this mapping procedure,
the $(k_{in}, k_{ou})$-core percolation can be well-defined
on the new graph with only directed arcs.
As a special example,
the $K$-core percolation on an undirected network
can be considered as
the $(K, K)$-core percolation on its directed network counterpart
after the above mapping.
In such a general setting,
we can consider the $(k_{in}, k_{ou})$-core percolation problem
on a graph $G = \{V, A^{un}, A^{di}\}$
with a node set $V$,
a set of undirected edges $A^{un}$,
and a set of directed arcs $A^{di}$.
The degree distribution $P(k_{u}, k_{+}, k_{-})$
for the graph $G$
can be defined as the probability of a randomly chosen node with
$k_{u}$ undirected interacted neighbors,
$k_{+}$ in-coming arcs,
and $k_{-}$ out-going arcs.
A special instance with mixed connections
can be generated from an undirected graph
by assigning a fraction $\rho$ of randomly chosen edges
into directed arcs,
in which an undirected edge between a node pair $i$ and $j$
is annotated as $(i, j)$ or $(j, i)$ with equal probabilities.

Yet, in order to keep our focus on the analytical solutions
and avoid the extra complexity from mixed types of connections,
here we consider the $(k_{in}, k_{ou})$-core percolation problem on directed networks
with only unidirectional connections among nodes,
or at most one directed arc between any two nodes.

\section{Theory}
\label{sec:theory}

The $(k_{in}, k_{ou})$-pruning process
can be applied on any network instance to reveal its core structure.
Yet for the directed uncorrelated random graphs,
we can derive a mean-field theory
based on the cavity method
\cite{
Mezard.Montanari-2009}
to theoretically predict the relative sizes of their $(k_{in}, k_{ou})$-cores.
With the cavity method,
we arrive at a set of self-consistent belief-propagation (BP) equations of cavity probabilities or messages,
whose formalism has roots in both statistical mechanics
\cite{
Bethe-ProcRSocLondA-1935}
and computer science
\cite{
Kschischang.Frey.Loeliger-IEEETransInfTheor-2001,
Yedidia.Freeman.Weiss-IEEETransInfTheor-2005}.
With the stable solutions of the BP equations,
we can calculate the quantities related to the core structure.

For a directed random graph $D = \{V, A\}$,
we define two cavity probabilities
to derive the BP equations.
On a randomly chosen arc $(i, j)$,
we start from the node $j$ which is in the core structure,
and arrive at the node $i$
following the opposite arc direction;
we define $\alpha$ as the probability
that $i$ is in the core structure
while $j$ is not considered.
In a similar sense,
we start from the node $i$ which is in the core structure,
and arrive at the node $j$
following the arc direction;
we define $\beta$ as the probability
that $j$ is in the core structure
while $i$ is not considered.
We further assume that
on an infinitely large random graph
which is uncorrelated and sparse,
a randomly chosen node $i$
has a locally tree-like structure,
or its neighbors show no correlation in states in the pruning process
after $i$ is removed.
With this approximation
\cite{
Mezard.Montanari-2009},
on a directed random graph with a degree distribution $P(k_{+}, k_{-})$,
we have the self-consistent equations for $\alpha$ and $\beta$.

%
\begin{eqnarray}
\label{eq:alpha}
\alpha
& = &
p \sum _{k_{+}, k_{-}} Q_{-}(k_{+}, k_{-})
[\sum _{k_{1} = k_{in}}^{k_{+}} 
\left( \begin{array}{c} k_{+} \\ k_{1} \end{array} \right)
\alpha ^{k_{1}} (1 - \alpha) ^{k_{+} - k_{1}}]
[\sum _{k_{2} = k_{ou} - 1}^{k_{-} - 1} 
\left( \begin{array}{c} k_{-} - 1 \\ k_{2} \end{array} \right)
\beta ^{k_{2}} (1 - \beta) ^{k_{-} - 1 - k_{2}}], \\
\label{eq:beta}
\beta
& = &
p \sum _{k_{+}, k_{-}} Q_{+}(k_{+}, k_{-})
[\sum _{k_{1} = k_{in} - 1}^{k_{+} - 1} 
\left( \begin{array}{c} k_{+} - 1 \\ k_{1} \end{array} \right)
\alpha ^{k_{1}} (1 - \alpha) ^{k_{+} - 1 - k_{1}}]
[\sum _{k_{2} = k_{ou}}^{k_{-}} 
\left( \begin{array}{c} k_{-} \\ k_{2} \end{array} \right)
\beta ^{k_{2}} (1 - \beta) ^{k_{-} - k_{2}}].
\end{eqnarray}
%
We briefly explain the derivation of the self-consistent equation of $\alpha$,
and the equation for $\beta$ follows a rather similar logic.
On a randomly chosen arc $(i, j)$ in a directed graph $D$,
from node $j$ following the opposite arc direction to node $i$,
the node $i$ has a probability $p$
that it remains in the residual graph after the initial removal step.
In order to further remain in the core structure after the pruning process,
the node $i$ should have both
at least $k_{in}$ in-neighbors in the core structure
and $k_{ou} - 1$ out-neighbors in the core structure
besides the node $j$
which is already in the core structure.
With the stable solutions of $\alpha$ and $\beta$,
the relative size of nodes in the core structure $n_{core}$
can be calculated.

\begin{equation}
\label{eq:n_core}
n_{core}
=
p \sum _{k_{+}, k_{-}} P(k_{+}, k_{-})
[\sum _{k_{1} = k_{in} }^{k_{+}} 
\left( \begin{array}{c} k_{+} \\ k_{1} \end{array} \right)
\alpha ^{k_{1}} (1 - \alpha) ^{k_{+} - k_{1}}]
[\sum _{k_{2} = k_{ou}}^{k_{-}} 
\left( \begin{array}{c} k_{-} \\ k_{2} \end{array} \right)
\beta ^{k_{2}} (1 - \beta) ^{k_{-} - k_{2}}].
\end{equation}
%
Here is an explanation of the equation for $n_{core}$:
if a newly added node can be in the final core structure,
it should have both at least $k_{in}$ in-neighbors and $k_{ou}$ out-neighbors
in the core structure,
provided that it is not removed in the initial step.

We can also derive the arc density $c_{core}$ of the core structure as

\begin{equation}
\label{eq:c_core}
c_{core}
= \sum _{k_{1} \geq k_{in}, k_{2} \geq k_{ou}} k_{1} P_{core}(k_{1}, k_{2})
= \sum _{k_{1} \geq k_{in}, k_{2} \geq k_{ou}} k_{2} P_{core} (k_{1}, k_{2}),
\end{equation}
while $P_{core} (k_{1}, k_{2})$
with $k_{1} \geq k_{in}$ and $k_{2} \geq k_{ou}$
denotes the degree distribution of the core structure 
and has the form

\begin{equation}
\label{eq:p_k1_k2}
P_{core} (k_{1}, k_{2})
=
\frac {1}{n_{core}}
p \sum _{k_{+} \geq k_{1}, k_{-} \geq k_{2}}
P(k_{+}, k_{-})
[\left( \begin{array}{c} k_{+} \\ k_{1} \end{array} \right)
\alpha ^{k_{1}}
(1 - \alpha)^{k_{+} - k_{-}}]
[\left( \begin{array}{c} k_{-} \\ k_{2} \end{array} \right)
\beta ^{k_{2}}
(1 - \beta)^{k_{-} - k_{2}}].
\end{equation}
%
We briefly explain Eq.\ref{eq:p_k1_k2} and then Eq.\ref{eq:c_core}.
If a node survives the initial removal
and has an in-degree $k_{1}$ and an out-degree $k_{2}$ after the pruning process,
surely it has $k_{1} \geq k_{in}$ and $k_{2} \geq k_{ou}$,
and it should have exactly $k_{1}$ in-neighbors and $k_{2}$ out-neighbors
in the core structure,
before the node is added into the original graph.
The above probability is divided by $n_{core}$
to derive the probability of the node
with $k_{1}$ in-neighbors and $k_{2}$ out-neighbors
in the core structure.
Thus we arrive at Eq.\ref{eq:p_k1_k2}.
After the average on all the in-degrees or all the out-degrees
in the core structure,
we have the mean arc density of the core structure
as showed in Eq.\ref{eq:c_core}.

Here we discuss the numerical method to
derive the stable solutions of $(\alpha, \beta)$
given $P(k_{+}, k_{-})$ and $p$,
with which we can calculate $n_{core}$ and $c_{core}$.
For the ease of discussion,
we denote the right-hand side of Eq.\ref{eq:alpha} as $f(\alpha, \beta)$
while a function $F(\alpha, \beta)$ is defined as
$F(\alpha, \beta) \equiv - \alpha + f(\alpha, \beta)$,
the right-hand size of Eq.\ref{eq:beta} as $g(\alpha, \beta)$
while a function $G(\alpha, \beta)$ is defined
as $G(\alpha, \beta) \equiv - \beta + g(\alpha, \beta)$.
For an $\alpha \in [0, 1]$,
we can calculate the corresponding stable $\beta$ with Eq.\ref{eq:beta}.
The calculation of stable $\beta$ goes like this:
for a $\beta \in [0, 1]$ given $\alpha$,
we can calculate $G(\alpha, \beta)$ with Eq.\ref{eq:beta};
with an incremental procedure with small steps for $\beta$,
we can find all the fixed $\beta$ in the range when $G(\alpha, \beta)$ change its sign;
the stable $\beta$ is the largest one among its fixed solutions.
With the stable $\beta$ given $\alpha$,
we can calculate $F(\alpha, \beta)$ with Eq.\ref{eq:alpha}.
With a similar procedure for calculating the stable $\beta$ with given $\alpha$,
we can calculate all the fixed $\alpha$,
thus the stable $\alpha$.
With the stable $\alpha$,
we can derive its corresponding stable $\beta$ with Eq.\ref{eq:beta}.
Thus we have the stable solution for $(\alpha, \beta)$.

We should mention that there are some special examples of
the $(k_{in}, k_{ou})$-core percolation.
When $k_{in} = \{0, 1\}$,
the term from the in-neighbors on the right-hand side of Eq.\ref{eq:beta} reduces to $1$;
when $k_{ou} = \{0, 1\}$,
the term from the out-neighbors on the right-hand side of Eq.\ref{eq:alpha} reduces to $1$.
In the case of $(k_{in}, k_{ou}) = (1, 0)$,
with the substitution of $1 - \alpha \rightarrow x$,
we arrive at the equation for the in-components of directed graphs;
in the same sense for the case of $(k_{in}, k_{ou}) = (0, 1)$
with the mapping $1 - \beta \rightarrow y$,
we have the theory for the out-components of directed graphs
\cite{
Newman.Strogatz.Watts-PRE-2001}.
In the case of $(k_{in}, k_{ou}) = (1, 1)$,
we further have the theory for the strongly connected components (SCC)
as the intersection of in-components and out-components in directed random graphs
\cite{
Tarjan-SIAMJourComp-1972,
Dorogovtsev.Mendes.Samukhin-PRE-2001,
Schwartz.Cohen.benAvarham.Barabasi.Havlin-PRE-2002}.
Transition behaviors in the above three cases of the percolation problem
in directed random graphs are continuous.
Yet for the cases with $k_{in} \geq 2$ or $k_{ou} \geq 2$,
later we will see that a quite different scenario of percolation transitions happens.

\section{Results}
\label{sec:results}

\subsection{Random networks}

\begin{figure}
\begin{center}
\includegraphics[width = 0.65 \linewidth]{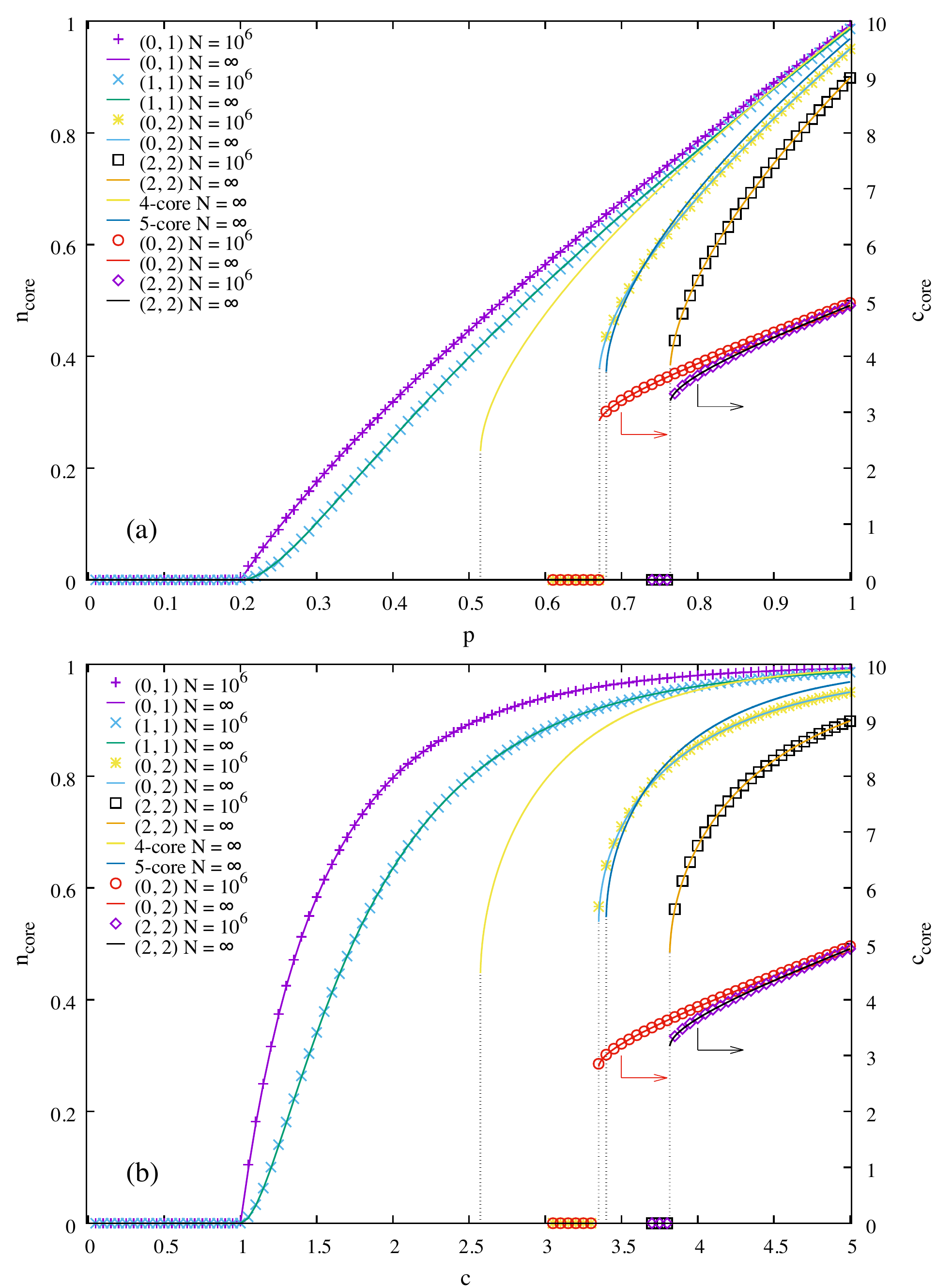}
\end{center}
\caption{
\label{fig:kkcore_er}
Normalized sizes of $(k_{in}, k_{ou})$-cores on directed ER random graphs.
We calculate the node fractions of the $(k_{in}, k_{ou})$-cores
from simulation and analytical theory
on directed ER random graphs.
The arc densities for $(0, 2)$- and $(2, 2)$-cores
are also derived and indicated by the right vertical y-axis.
In (a),
we consider ER graphs with an arc density $c = 5.0$
with different initial fraction $p$.
In (b),
we consider ER graphs with different arc density $c$
with an initial fraction $p = 1.0$.
Points are for the simulation results,
each of which is derived from a pruning process
on a single directed ER random graph instance with a node size $N = 10^6$.
Solid lines are for the analytical results in the case of graphs with a node size $N = \infty$.
Dashed lines indicate the discontinuous transitions from the analytical theory.
As a comparison,
we also show the analytical results of the $4$- and $5$-cores
on infinitely large undirected ER random graphs in both (a) and (b),
whose notation of the arc density $c$ corresponds to the edge-node ratio,
or half the the mean degree or the mean connectivity of these undirected graphs.}
\end{figure}
\begin{figure}
\begin{center}
 \includegraphics[width = 0.90 \linewidth]{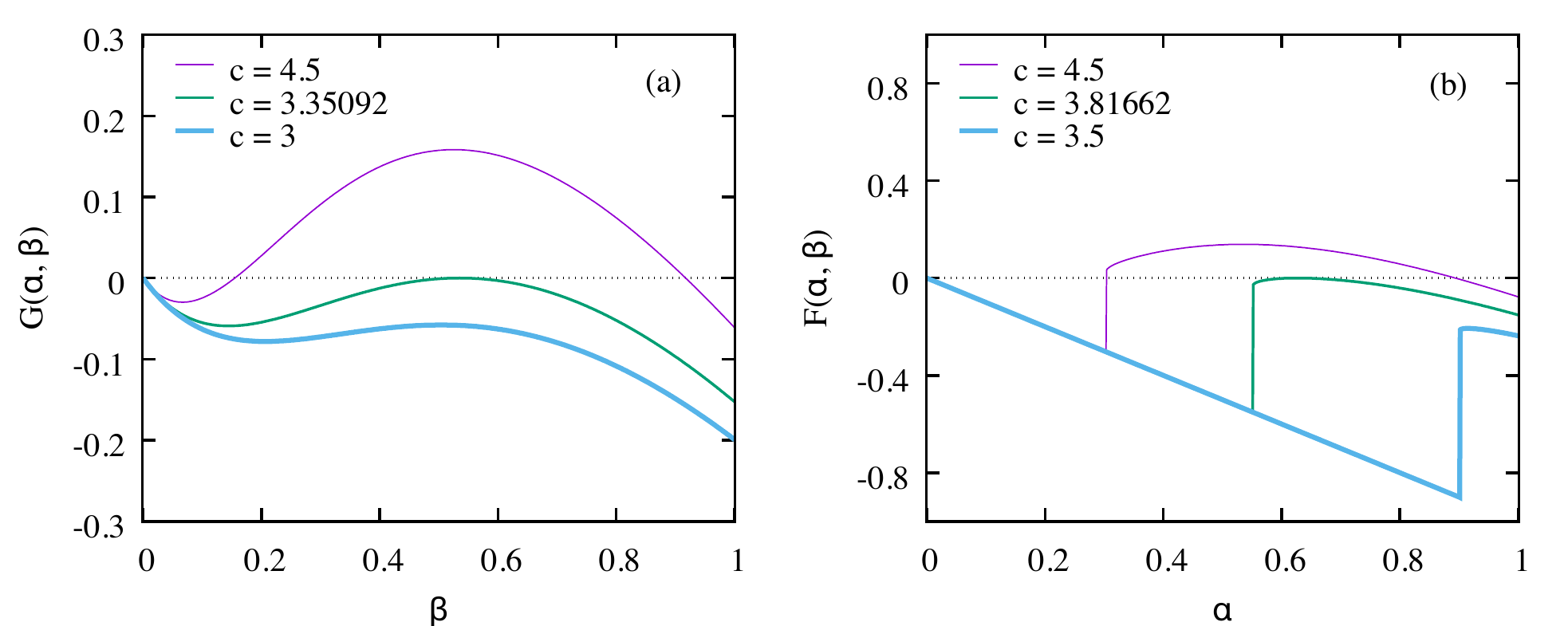}
\end{center}
\caption{
 \label{fig:kkcore_fb_fa_er}
Transition behaviors of stable solutions of $\alpha$ and $\beta$ for $(k_{in}, k_{ou})$-core percolation
on directed ER random graphs.
The fixed solutions of $\beta$ for $(0, 2)$-core percolation in (a)
are calculated as $G(\alpha, \beta) = 0$ and
the fixed solutions of $\alpha$ for $(2, 2)$-core percolation in (b)
are calculated as $F(\alpha, \beta) = 0$
with the analytical theory 
on infinitely large directed ER random graphs with an initial fraction $p = 1.0$.
In (a),
when the arc density $c = c^{*}$ with $c^{*} \approx 3.35092$,
a second fixed and also the stable solution of $\beta$
shows up
besides the trivial fixed solution $\beta = 0$,
thus both the stable $\beta$ and $n_{core}$ experience a sudden jump.
In (b),
a similar transition behavior of stable $\alpha$
happens at $c = c^{*}$ with $c^{*} \approx 3.81662$.}
\end{figure}
\begin{figure}
\begin{center}
 \includegraphics[width = 0.65 \linewidth]{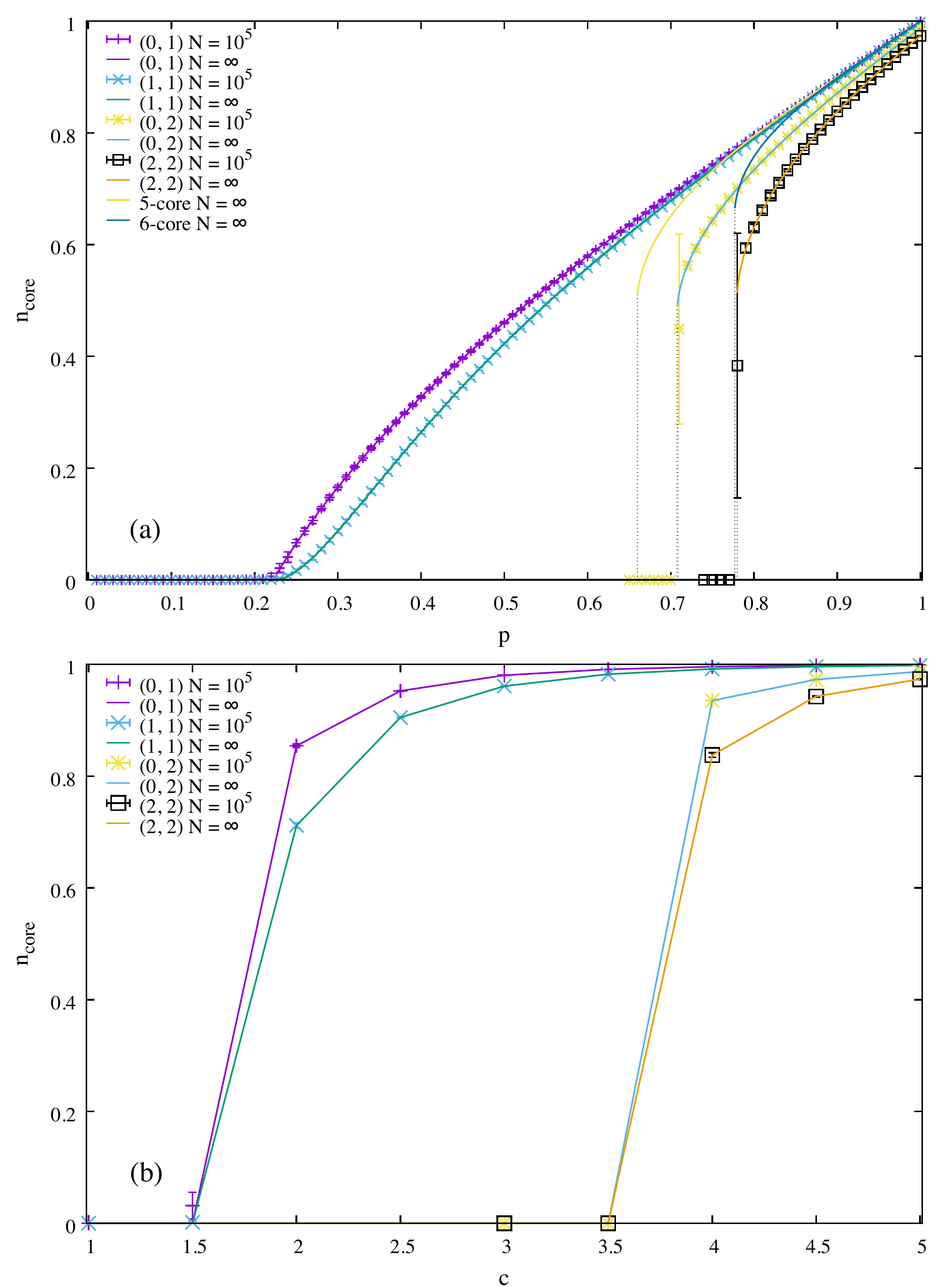}
\end{center}
\caption{
\label{fig:kkcore_rr}
Normalized sizes of $(k_{in}, k_{ou})$-cores on directed RR graphs.
The node fractions of some $(k_{in}, k_{ou})$-cores
are calculated from the simulation and the analytical theory.
In (a),
the node fractions are calculated
on directed RR graphs with an arc density $c = 5.0$
with different initial fraction $p$.
For a comparison,
the relative sizes of $5$- and $6$-cores are also calculated on infinitely large
undirected RR graphs with a degree $k_{0} = 10$ with the analytical theory.
In (b),
the node fractions are calculated
on directed RR graphs with different arc densities
with an initial fraction $p = 1$.
Points are for the averaged results of simulation
on $40$ independently generated graph instances with a node size $N = 10^5$.
The standard deviation for each data point from simulation is also shown.
In (a),
the solid lines are for the analytical results on infinitely large graphs,
while the dashed lines are for the discontinuous transitions from the analytical theory.
In (b),
the points of intersection between solid line segments are for the analytical results on infinitely large graphs.}
\end{figure}
\begin{figure}
\begin{center}
 \includegraphics[width = 0.65 \linewidth]{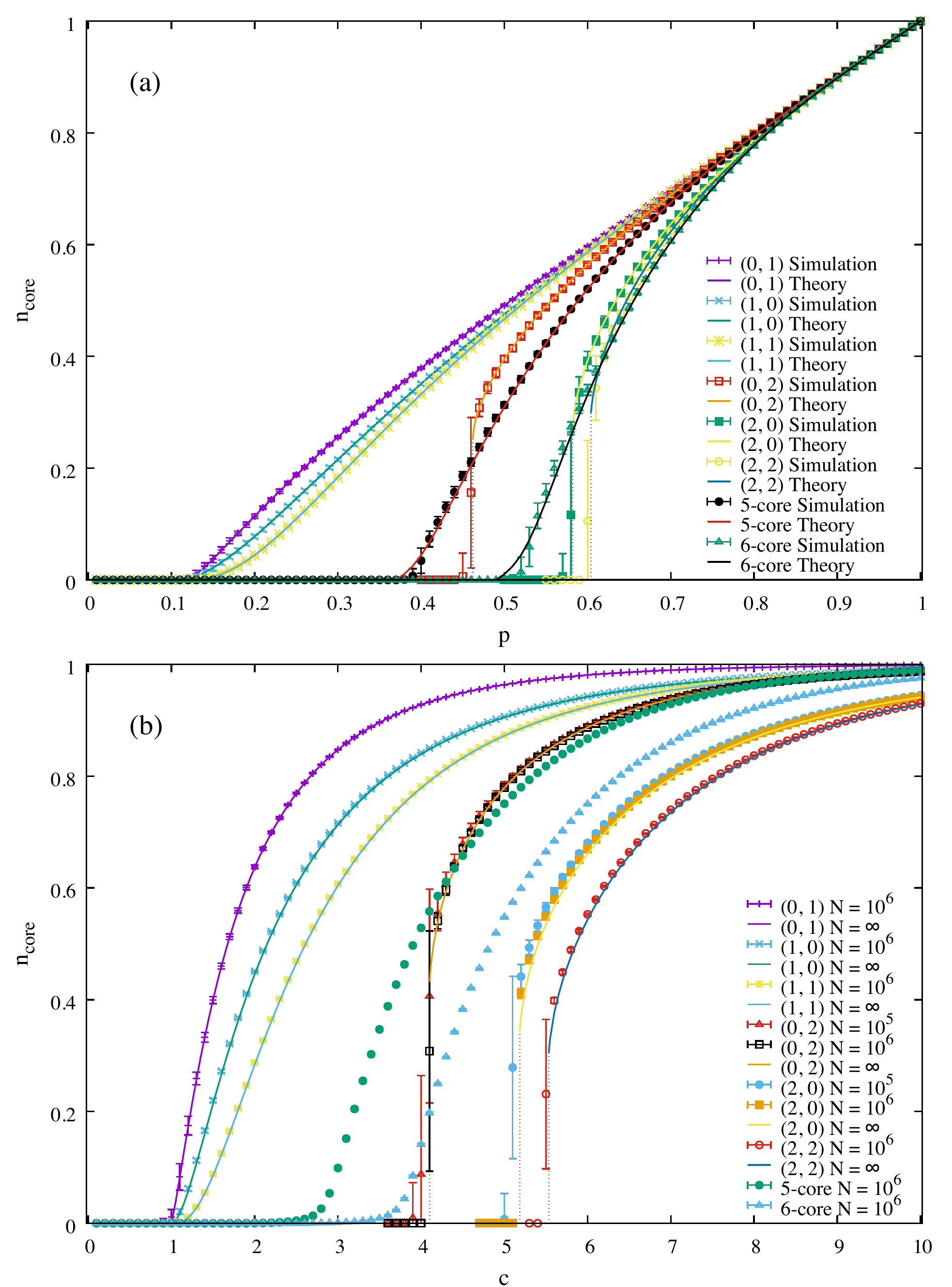}
\end{center}
\caption{\label{fig:kkcore_sf_sm}
Normalized sizes of $(k_{in}, k_{ou})$-cores of directed SF networks.
In (a),
we calculate the node fractions of the $(k_{in}, k_{ou})$-cores
from the simulation and the analytical theory
on a directed SF network instance generated by the configurational model.
The SF network instance
has a node size $N = 10^5$ with
an in-degree exponent $\gamma _{+} = 2.5$,
an out-degree exponent $\gamma _{-} = 3.0$,
and a minimal degree $k_{+}^{min} = k_{-}^{min} = 4$
and a maximal degree $k_{+}^{max} = k_{-}^{max} = \sqrt N$
for both in-degrees and out-degrees.
We also calculate the relative sizes of
the $5$- and $6$-cores
with simulation and analytical theory
on the undirected network counterpart
after ignoring the arc directions.
Points are for the simulation results
averaged from those on $40$ independently generated initial configurations with a given initial fraction $p$,
while their standard deviations are also shown.
Solid lines are for the analytical results
based on the empirical degree distribution of the graph instance.
Dashed lines indicate the abrupt transitions by the analytical theory.
In (b),
we calculate the node fractions of the $(k_{in}, k_{ou})$-core structures
from the simulation and the analytical theory
on asymptotical SF networks generated with the static model.
The SF networks have
an in-degree exponent $\gamma _{+} = 2.5$ and
an out-degree exponent $\gamma _{-} = 3.0$.
We also calculate the relative sizes of
the $5$- and $6$-cores
with simulation on the undirected network counterparts.
Each data point is for the simulation result averaged on
$40$ independently generated directed SF instances with the node size $N = 10^6$ or $10^5$
as indicated in the legend.
Standard deviations for all the simulation results are presented.
Solid lines are for the analytical results on infinitely large graphs.
Dashed lines indicate the discontinuous transitions from the analytical theory.}
\end{figure}
%

Here we consider the $(k_{in}, k_{ou})$-cores of some model directed random networks.
We leave the simplified equations for the $(k_{in}, k_{ou})$-core percolation
on directed random networks in Appendix A.
In the paper,
we also compare the $(k_{in}, k_{ou})$-cores
with the $K$-cores on the undirected version of directed random networks,
and we left the theory of the $K$-core percolation problem
on undirected networks in Appendix B.

First we consider the case of directed Erd\"{o}s-R\'{e}nyi (ER) random graphs
\cite{
Erdos.Renyi-PublMath-1959,
Erdos.Renyi-Hungary-1960}.
To generate a directed ER random graph,
we first generate an undirected ER random graph instance;
then for each undirected edge between a node pair $i$ and $j$,
we assign a direction to form a directed arc $(i, j)$ or $(j, i)$ with an equal probability.
For a directed ER random graph with an arc density $c$,
the degree distribution is $P(k_{+}, k_{-}) = e^{-c}c^{k_{+}}/k_{+}! \times e^{-c}c^{k_{-}}/k_{-}!$.
In figure \ref{fig:kkcore_er},
we show the normalized sizes of $(k_{in}, k_{ou})$-cores
with different initial fraction $p$ and arc density $c$.
We can see that for $k_{in}, k_{ou} = \{0, 1\}$,
the emergence of the core structure undergoes a continuous transition.
While in the cases with $k_{in} \ge 2$ or $k_{ou} \ge 2$,
the emergence shows a first-order transition.
This constitutes our first interesting result,
as it is well-known that the $K$-core percolation with $K \geq 3$
on undirected ER graphs is discontinuous.
An intuitive understanding of this result is that
a node removal procedure,
in which part of the neighbors of a node
rather than all its residual neighbors
can result in the failure of the node,
usually leads to a more aggressive network breakdown process.
A similar logic shows
in the model with an extra inducing effect from the failed nodes
on undirected networks
\cite{
Zhao.Zhou.Liu-NatCommun-2013}
and the models with failure propagation through coupling or interdependency between nodes
on multilayer structures
\cite{
Buldyrev.etal-Nature-2010,
AzimiTafreshi.etal-PRE-2014}.
A theoretical explanation of the discontinuous emergence of core structures
can be understood from the behaviors of the stable solutions
of $\alpha$ and $\beta$.
We first take the $(0, 2)$-core percolation problem as an example.
For the $(0, 2)$-core percolation with $p = 1$,
the iterative equations and the normalized core structure are
$\alpha = 1 - e^{- c \beta},  \beta = 1 - e^{- c \beta} (1 + c \beta), n_{core} = \beta$.
Since it only depends on $\beta$,
the transition behavior of $n_{core}$
can be explained by the stable $\beta$.
With the functions defined in the previous section,
we show $G(\alpha, \beta)$
in figure \ref{fig:kkcore_fb_fa_er} (a)
as we can find all the fixed $\beta$ with $G(\alpha, \beta) = 0$.
For $c < c^{*}$
as the critical arc density $c^{*} \approx 3.35092$,
there is only one fixed and also the stable solution $\beta = 0$,
correspondingly $n_{core} = 0$.
When $c = c^{*}$,
a second fixed and also the stable solution
$\beta \approx 0.535$
shows up abruptly,
thus a sudden emergence of the core structure with
a critical normalized size $n_{core}^{*} = \beta$.
As $c > c^{*}$,
the stable solution of $\beta$
and the corresponding $n_{core}$
increase further.
For the case of $(2, 2)$-core percolation,
$n_{core}$ depends on both the stable solutions of $\alpha$  and $\beta$.
In figure \ref{fig:kkcore_fb_fa_er} (b),
a rather similar transition pattern happens for the stable $\alpha$
and correspondingly the stable $\beta$.
The analysis of a sudden emergence of nontrivial stable solutions of $\alpha$ and $\beta$
and the corresponding abrupt behavior of $n_{core}$ on infinitely large random graphs
can be applied to other cases of
$(k_{in}, k_{ou})$-core percolation
with $k_{in} \geq 2$ or $k_{ou} \geq 2$.
In Sec.\ref{sec:hybrid_transitions},
we give a general proof of this discontinuity,
where we also show that in the supercritical region of a discontinuous transition
of the core structure,
a scaling property applies as $n_{core} - n_{core}^{*} \propto (c - c^{*})^{1/2}$
with $\delta c \equiv c - c^{*} \ll 1$
which is independent of the network types
and the model parameters.
This type of hybrid transition
can be found in various percolation problems
on different network types such as
\cite{
Dorogovtsev.Goltsev.Mendes-PRL-2006,
Zhao.Zhou.Liu-NatCommun-2013,
AzimiTafreshi.etal-PRE-2014,
Liu.Csoka.Zhou.Posfai-PRL-2012,
Liu.etal-NatCommun-2017}.

The second interesting result on directed ER random graphs
is that a $(0, 2)$-core emerges around an arc density
where a $5$-core suddenly shows up on the undirected counterparts.
As in figure \ref{fig:kkcore_er} (a),
a $5$-core on an undirected ER ensemble with a mean degree $d = 10.0$
emerges at an initial fraction $p \approx 0.680$,
while a $(0, 2)$-core on the directed ER random graph ensemble with an arc density $c \equiv d / 2 = 5.0$
shows up at an initial fraction $p \approx 0.671$;
in figure \ref{fig:kkcore_er} (b),
a $5$-core on an undirected ER ensemble
shows up at a mean degree $d \approx 6.800$,
correspondingly an edge-node ratio $3.400$,
while a $(0, 2)$-core on the directed ER random graph ensemble
emerges at an arc density $c \approx 3.351$.
The above comparision means that for a directed ER random graph
which permits a certain $(k_{in}, k_{ou})$-core,
a macroscopic $K$-core with a much larger parameter $K$ than $k_{in}$ and $k_{ou}$ is possible
when the arc directions are ignored.
It is reasonable to expect that the $(k_{in}, k_{ou})$-core pruning process
is generally a more aggressive network breakdown procedure than
the $K$-core pruning process after ignoring arc directions,
for example in the case of $k_{in} + k_{ou} = K$,
yet here we provide an analytical ground for this intuitive understanding.

We then consider the percolation problem on the directed regular random (RR) graphs.
We generate a directed RR graph instance from an undirected RR graph instance,
in which each edge of the undirected instance is assigned randomly with a direction
just like we do in the generation of directed ER random graphs.
For an undirected RR graph with an integer degree $k_{0}$
($k_{0}$ undirected neighbors for each node),
a corresponding directed RR graph instance
with an arc density $c \equiv k_{0}/2$
can be generated.
In figure \ref{fig:kkcore_rr},
we show the result for the normalized core structures on directed RR graphs.

We further consider the percolation problem
on the scale-free (SF) networks.
SF networks show power-law degree distributions,
and are ubiquitous in the real world
\cite{
Barabasi.Albert-Science-1999}.
We consider the case of SF networks
with a form of degree distribution without degree correlation as
$P(k_{+}, k_{-}) \approx P_{+}(k_{+}) P_{-}(k_{-})$
with $P_{\pm}(k_{\pm}) \propto k_{\pm}^{- \gamma _{\pm}}$,
while $\gamma _{+}$ and $\gamma _{-}$
are respectively the exponents for the in-degrees and the out-degrees.

We first consider the directed SF networks
 generated with the configurational model
 \cite{
 Newman.Strogatz.Watts-PRE-2001,
 Zhou.Lipowsky-PNAS-2005}
which are constructed based on sequences of in-degrees and out-degrees
generated directly from the power-law distribution.
To construct such a directed SF network with a node size $N$
with the degree distribution
$P(k_{+}, k_{-}) \propto k_{+}^{- \gamma _{+}} k_{-}^{- \gamma _{-}}$,
we need to further specify
a minimal degree $k^{min}_{+}$ and a maximal degree $k^{max}_{+}$ for in-degrees,
and a minimal degree $k^{min}_{-}$ and a maximal degree $k^{max}_{-}$ for out-degrees.
With the parameters $(\gamma _{+}, k^{min}_{+}, k^{max}_{+})$,
we can construct an in-degree sequence with $P_{+}(k_{+}) \propto k_{+}^{- \gamma _{+}}$
with $k_{+}^{min} \leq k_{+} \leq k_{+}^{max}$,
while there are $N P_{+}(k_{+})$ nodes each with $k_{+}$ in-coming half-arcs
and in total $E_{+} (= \sum _{k_{+}} k_{+} N P_{+}(k_{+}))$ in-coming half-arcs.
The same procedure is adopted to construct an out-degree sequence
through $P_{-}(k_{-}) \propto k_{-}^{- \gamma _{-}}$
with $k_{-}^{min} \leq k_{-} \leq k_{-}^{max}$,
while there are $N P_{-}(k_{-})$ nodes each with $k_{-}$ out-going half-arcs
and in total $E_{-} (= \sum _{k_{-}} k_{-} N P _{-}(k_{-}))$ out-going half-arcs.
We further make sure that the numbers of in-coming and out-going half-arcs are equal
as $E = (E_{+} + E_{-}) / 2$
by removing and adding in-coming and out-going half-arcs correspondingly.
Then an in-coming half-arc and an out-going half-arc
are randomly chosen and paired to establish a genuine directed arc
until there is no half-arc left.
The graph instance generated with this procedure
has a degree distribution
$P(k_{+}, k_{-}) \approx
k_{+}^{- \gamma _{+}}/\sum _{k_{+}} k_{+}^{- \gamma _{+}} \times
k_{-}^{- \gamma _{-}}/\sum _{k_{-}} k_{-}^{- \gamma _{-}}$
and an arc density
$c = \sum _{k_{+}, k_{-}} k_{+} P(k_{+}, k_{-}) = \sum _{k_{+}, k_{-}} k_{-} P(k_{+}, k_{-})$,
while in the summations
$k_{+}^{min} \leq k_{+} \leq k_{+}^{max}$ and $k_{-}^{min} \leq k_{-} \leq k_{-}^{max}$.
In figure \ref{fig:kkcore_sf_sm} (a),
on a SF network instance with an in-degree exponent $\gamma _{+} = 2.5$
and an out-degree exponent $\gamma _{-} = 3.0$,
we show the sizes of core structures
from both simulation and analytical theory
with different initial fraction $p$.
Our analytical theory predicts well the birth points and the relative sizes of core structures
on the graph instance even with a finite size.

Directed SF networks can also be generated with the static model
\cite{
Goh.Kahng.Kim-PRL-2001,
Catanzaro.PastorSatorras-EPJB-2009}.
For a SF network instance
with an in-degree exponent $\gamma _{+}$
and an out-degree exponent $\gamma _{-}$,
the construction procedure
goes like this:
for an empty graph with $N$ nodes indexed as $i \in \{1, 2, .., N\}$,
each node is assigned with an in-degree weight
$w_{+}^{i} \propto i^{- \xi_{+}}$ and
an out-degree weight
$w_{-}^{i} \propto i^{- \xi_{-}}$
as $\xi _{\pm} \equiv 1 / (\gamma _{\pm} - 1)$;
in the arc establishment process,
two distinct nodes, say nodes $i$ and $j$,
are chosen proportionally to their weights $w_{+}^{i}$ and $w_{-}^{j}$, respectively,
and are connected into a directed arc $(i, j)$;
with this process, we establish $M = c N$ arcs with an arc density $c$ in the graph.
Based on the theory in
\cite{
Goh.Kahng.Kim-PRL-2001,
Catanzaro.PastorSatorras-EPJB-2009},
the graph generated with this method has
a degree distribution $P(k_{+}, k_{-}) = P(k_{+}) P(k_{-})$ as
$P(k_{\pm})
=
\frac {1}{\xi _{\pm}}
\frac {(c (1 - \xi _{\pm}) )^{k_{\pm}}}{k_{\pm}!}
\int _{1}^{\infty}
\mathrm{d} t e^{- c (1 - \xi _{\pm}) t} t^{k_{\pm} - 1 - 1 / \xi _{\pm}}$.
With large $k_{+}, k_{-}$,
we have $P(k_{+}, k_{-}) \propto k_{+}^{- \gamma _{+}} k_{-}^{- \gamma _{-}}$.
In figure \ref{fig:kkcore_sf_sm} (b),
we calculate the core structures on approximate SF networks generated by the static model with
an in-degree exponent $\gamma _{+} = 2.5$
and an out-degree exponent $\gamma _{-} = 3.0$.
We can see that for $(0, 2)$-cores
which are revealed by a pruning process based only on the out-degrees of nodes,
results show quite small differences
between the simulation results on instances with different node sizes
and the analytical theory on infinitely large graphs.
While thing is different in the case of $(2, 0)$-cores.
It is a well-known observation
that when the degree exponent $\gamma \geq 3.0$,
an undirected SF graph generated with the static model
is becoming more like an ER random graph
with an increasing degree exponent
\cite{
Goh.Kahng.Kim-PRL-2001,
Catanzaro.PastorSatorras-EPJB-2009}.
We can say that this observation still holds
in the pruning processes based on only $k_{in}$ or $k_{ou}$
for respectively the in- or out-degrees
of the directed graphs generated with the static model.

From the results on the directed RR and SF networks,
we can see again the discontinuity with $k_{in} \geq 2$ or $k_{ou} \geq 2$
and a relatively large arc density for the discontinuous emergence of $(k_{in}, k_{ou})$-cores
compared with the $K$-cores on undirected network counterparts,
just like we see in the results on the directed ER random graphs.

Apart from the major observations between
the model parameters ($k_{in}$ and $k_{ou}$) and the core sizes,
we also have two observations related to degree distributions and the core sizes.
(1) For directed ER graphs, RR graphs, and SF networks with large enough degree exponents,
the analytical theory predicts nearly the exact relative sizes of the core structures
even for finite-size graphs
based only on their degree distributions,
which means that
the size of the $(k_{in}, k_{ou})$-core of a random graph
is much coded in its degree distribution.
(2) For networks with different distributions
of in-degrees and out-degrees
like SF networks we consider above,
$(k_{in}, k_{ou})$- and $(k_{ou}, k_{in})$-cores with $k_{in} \neq k_{ou}$
show a difference.
From these two aspects,
later we will discuss the core structures on real network data sets
which need a more comprehensive structural characterization
than the one for random networks.

\subsection{Real networks}

\begin{table*}
\caption{
  \label{tab:real_info}
  \textbf{Real directed networks.}
  For each real network,
  \textbf{Type} and Name
  list the general category and the name,
  Description
  a brief description,
  $N$
  the size of nodes,
  and $M$
  the size of directed arcs.}
\begin{center}
\begin{tabular}{llrrrrrrr}
\hline
\textbf{Type} and Name
& Description
& $N$
& $M$ \\
\hline
\textbf{Regulatory} \\
EGFR
\cite{Fiedler.Mochizuki.Kurosawa.Saito-JDynDiffEquat-2013a}
& Signal transduction network of EGF receptor.
& $61$
& $112$ \\
\textsl{E. coli}
\cite{Mangan.Alon-PNAS-2003}
& Transcriptional regulatory network of \textsl{E. coli}.
& $418$
& $519$ \\
\textsl{S. cerevisiae}
\cite{Milo.etal-Science-2002}
& Transcriptional regulatory network of \textsl{S. cerevisiae}.
& $688$
& $1,079$ \\
PPI 
\cite{Vinayagam.etal-ScienceSignaling-2011}
& Protein-protein interaction network of human.
& $6,339$
& $34,814$ \\
\hline
\textbf{Metabolic}\\
\textsl{C. elegans}
\cite{Jeong.Tombor.Albert.Oltvai.Barabasi-Nature-2000}
& Metabolic network of \textsl{C. elegans}.
& $1,469$
& $3,447$ \\
\textsl{S. cerevisiae}
\cite{Jeong.Tombor.Albert.Oltvai.Barabasi-Nature-2000}
& Metabolic network of \textsl{S. cerevisiae}.
& $1,511$
& $3,833$ \\
\textsl{E. coli}
\cite{Jeong.Tombor.Albert.Oltvai.Barabasi-Nature-2000}
& Metabolic network of \textsl{E. coli}.
& $2,275$
& $5,763$ \\
\hline
\textbf{Neuronal}\\
\textsl{C. elegans}
\cite{Watts.Strogatz-Nature-1998}
& Neuronal network of \textsl{C. elegans}.
& $297$
& $2,359$ \\
\hline
\textbf{Ecosystems}\\
Chesapeake
\cite{eco.Chesapeake-1989}
& Ecosystem in Chesapeake Bay.
& $39$
& $176$ \\
St. Marks 
\cite{eco.StMarks-1998}
& Ecosystem in St. Marks River Estuary.
& $54$
& $353$ \\
Florida
\cite{eco.Florida-1998}
& Ecosystem in Florida Bay.
& $128$
& $2,106$ \\
\hline
\textbf{Electric circuits}\\
s208
\cite{Milo.etal-Science-2002}
& Electronic sequential logic circuit.
& $122$
& $189$  \\
s420
\cite{Milo.etal-Science-2002}
& Same as above.
& $252$
& $399$ \\
s838
\cite{Milo.etal-Science-2002}
& Same as above.
& $512$
& $819$ \\
\hline
\textbf{Ownership}\\
USCorp
\cite{Norlen.Lucas.Gebbie.Chuang-ProdIntTeleSoc-2002}
& Ownership network of US corporations.
 & $7,253$
 & $6,724$ \\
\hline
\textbf{Internet p2p} \\
Gnutella04
\cite{
Leskovec.Kleinberg.Faloutsos-ACM-2007,
Ripeanu.Foster.Iamnitchi-ICompJour-2002}
 & Gnutella peer-to-peer file sharing network.
 & $10,876$ 
 & $39,994$ \\
 Gnutella30
\cite{
Leskovec.Kleinberg.Faloutsos-ACM-2007,
Ripeanu.Foster.Iamnitchi-ICompJour-2002}
 & Same as above (at different time).
 & $36,682$
 & $88,328$ \\
 Gnutella31
 \cite{
 Leskovec.Kleinberg.Faloutsos-ACM-2007,
Ripeanu.Foster.Iamnitchi-ICompJour-2002}
 & Same as above (at different time).
 & $62,586$
 & $147,892$ \\
\hline
\textbf{Social}\\
WiKi-Vote
 \cite{
Leskovec.Huttenlocher.Kleinberg-CHI-2010,
Leskovec.Huttenlocher.Kleinberg-WWW-2010}
 & Wikipedia who-votes-on-whom network.
 & $7,115$
 & $103,689$  \\
\hline
\hline
\end{tabular}
\end{center}
\end{table*}
\begin{figure}
\begin{center}
 \includegraphics[width = 0.85 \linewidth]{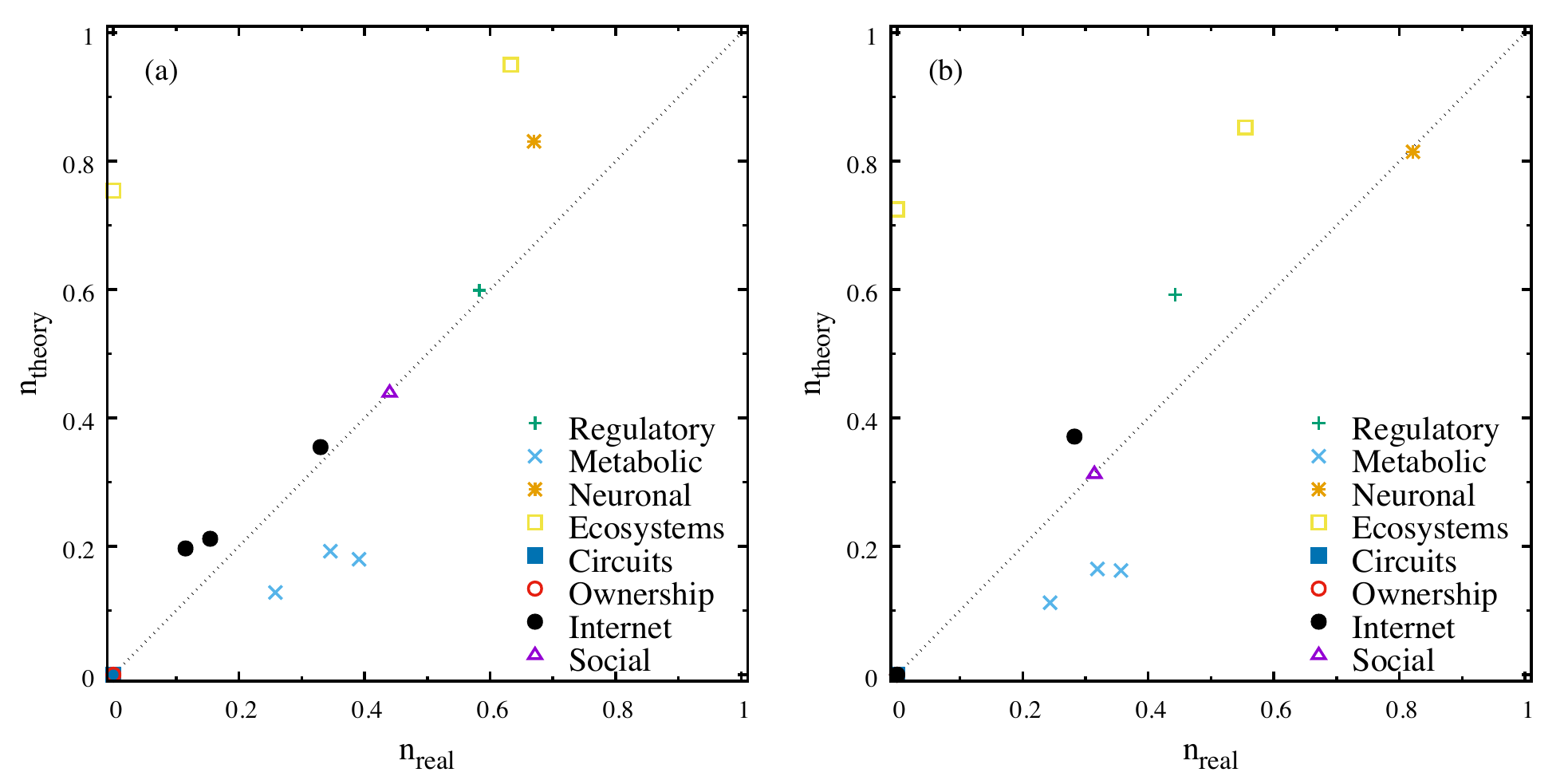}
\end{center}
\caption{\label{fig:kkcore_real}
Normalized sizes of $(k_{in}, k_{ou})$-cores on real network instances.
We calculate the relative sizes of
the $(0, 2)$-cores in (a)
and the $(2, 0)$-cores in (b)
for the $19$ real network instances
by the simulation ($n_{real}$) and
the analytical theory ($n_{theory}$)
with only their empirical degree distributions as inputs.
The initial fraction $p = 1.0$.}
\end{figure}
\begin{figure}
\begin{center}
 \includegraphics[width = 0.65 \linewidth]{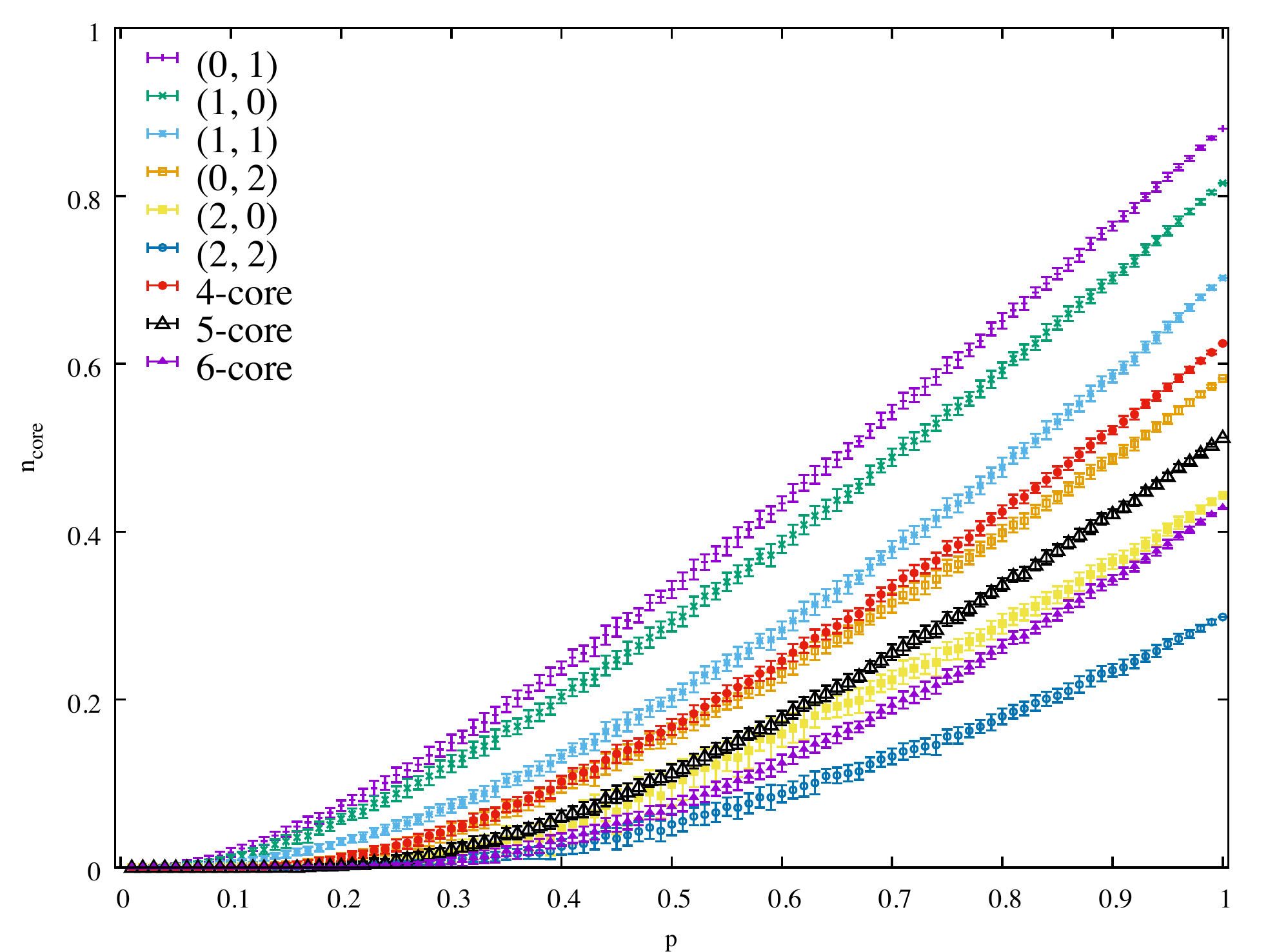}
\end{center}
\caption{\label{fig:kkcore_real_p}
Normalized sizes of $(k_{in}, k_{ou})$-cores
and $K$-cores of a protein-protein interaction network
against random node damages.
$K$-cores are derived on the undirected version of the network instance.
All the data points are averaged from the results
of pruning processes performed on $40$ independently generated configurations
with a given initial fraction $p$.
The standard deviation for each data point is also shown.}
\end{figure}
%

We consider here the $(k_{in}, k_{ou})$-cores on the real networks.
In Tab.\ref{tab:real_info}
we list the names, the node sizes, and the arc sizes
for the $19$ real network instances
we will consider.
For these networks,
self-connections are removed,
yet the multiple directional connections between nodes are permitted.

In figure \ref{fig:kkcore_real},
we show the sizes of the $(0, 2)$- and $(2, 0)$-cores
from the simulation on networks instances and
from the analytical theory based on the empirical degree distributions of network data sets.
We should mention that,
the analytical result derived with empirical degree distributions
is just like averaging the sizes of core structures
on instances when the in-neighbors and the out-neighbors for each node are fully randomized separately
(for example, by switching corresponding predecessors or successors of two randomly chosen arcs).
We can see that,
for most of these real networks,
the results from the analytical theory show a considerable discrepancy
from those derived from the simulation.
This discrepancy comes from a fact that
a sufficient description of a real network
is far from a degree distribution
without degree correlation.
Possible reasons are
the degree correlation
\cite{
Newman-PRL-2002}, 
the community structure
\cite{
Leicht.Newman-PRL-2008,
Malliaros.Vazirgiannis-PhysRep-2013},
the hierarchical structure
\cite{
CorominasMurtra.etal-PNAS-2013},
and so on.

In figure \ref{fig:kkcore_real_p},
for a protein-protein interaction network
\cite{Vinayagam.etal-ScienceSignaling-2011}
with a node size $N = 6,339$
and an arc size $M = 34,814$,
we show the shrinkage of the $(k_{in}, k_{ou})$-cores
and also the $K$-cores on its undirected network counterpart
against a random node damage process (see from right to left).
The random node damage is carried out through the initial fraction $p$.
This network shows a significant robustness against random node damages
in quite a range of $p$.
We can further see a clear difference between the sizes of
$(k_{in}, k_{ou})$- and $(k_{ou}, k_{in})$-cores
when $k_{in} \neq k_{ou}$.
It's reasonable to understand that the real networks are usually embedded with some information processing tasks,
in which the roles of in-coming and out-going arcs (directed interactions) have some intrinsic differences,
expressing themselves in the degree distributions and the higher-order structures.
The $(k_{in}, k_{ou})$-core pruning process
can thus be adopted as an intermediate method to reveal this structural subtlety
in real directed networks.

\subsection{Hybrid transitions}
\label{sec:hybrid_transitions}

\begin{figure}
\begin{center}
 \includegraphics[width = 0.90 \linewidth]{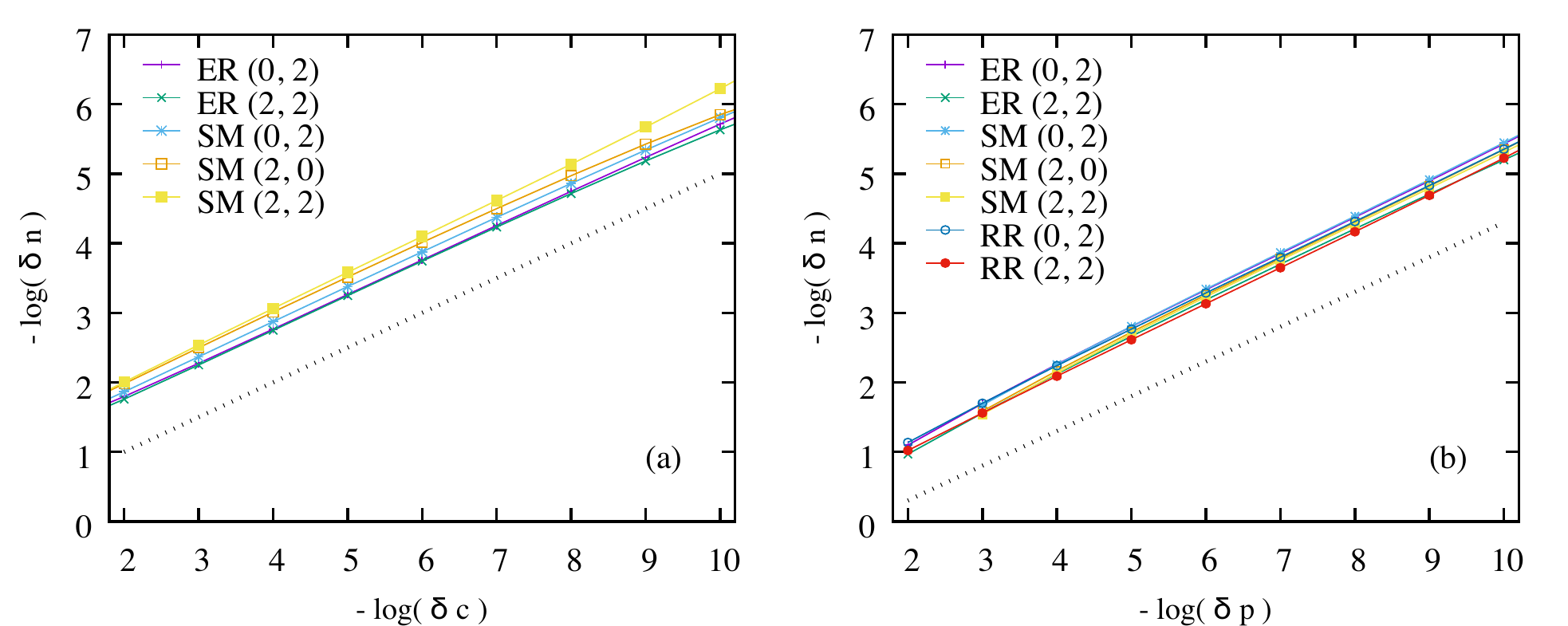}
\end{center}
\caption{\label{fig:kkcore_scaling}
Scaling property above the critical points
for the $(k_{in}, k_{ou})$-core percolation on directed random graphs.
In (a),
we show the differences of the relative core sizes $n_{core}$
and the arc densities $c$
respective to their critical values
for the directed ER random graphs (ER)
and the directed SF networks generated with the static model (SM)
both with the initial fraction $p = 1.0$.
In (b),
we show the differences of the relative core sizes $n_{core}$
and the initial fractions $p$
respective to their critical values
for the directed ER random graphs (ER) with an arc density $c = 5.0$,
the directed SF networks generated with the static model (SM) with an arc density $c = 6.0$,
and the directed RR graphs (RR) with an arc density $c = 5.0$.
The x-axis in (a) is rescaled as $-\log (\delta c)$ with $\delta c \equiv c - c^{*}$,
and the x-axis in (b) is rescaled as $-\log (\delta p)$ with $\delta p \equiv p - p^{*}$,
while the y-axes for both (a) and (b) are rescaled as
$-\log (\delta n)$ with $\delta n \equiv n_{core} - n_{core}^{*}$.
Points jointed by solid line segments 
are for the analytical results on infinitely large graphs.
Dashed lines
are both with a slope $1/2$ and
are simply for a comparison with data points.}
\end{figure}

Here we consider the discontinuity and the scaling behavior
relating to the hybrid transitions in
the $(k_{in}, k_{ou})$-core percolation problem.
We prove that no matter the network type,
the degree constraint parameters with $k_{in} \geq 2$ or $k_{ou} \geq 2$
lead to a discontinuous transition for the emergence of the core structures
on infinitely large graphs,
and the scaling property in the supercritical region of the hybrid transitions
has an exponent $1/2$.
Other hybrid transitions
can be found in the percolation problems as
\cite{
Dorogovtsev.Goltsev.Mendes-PRL-2006,
Zhao.Zhou.Liu-NatCommun-2013,
AzimiTafreshi.etal-PRE-2014,
Liu.Csoka.Zhou.Posfai-PRL-2012,
Liu.etal-NatCommun-2017}.

\subsubsection{Discontinuity}

We prove that the transition behavior
in the $(k_{in}, k_{ou})$-core percolation problem
is discontinuous
when $k_{in} \geq 2$ or $k_{ou} \geq 2$.
We focus on the case with $k_{in} \geq 2$ here,
and the case with $k_{ou} \geq 2$ has a similar analysis.

When $k_{in} \geq 2$,
the terms from the in-degrees on the right-hand sides of
Eqs.\ref{eq:alpha} and \ref{eq:beta}
both reduces to $0$ when $\alpha = 0$.
Thus $(\alpha, \beta) = (0, 0)$
is always a fixed solution.
Correspondingly,
there is a trivial core size as $n_{core} = 0$.
We then analyze how the nontrivial stable solutions
of $\alpha$ and $\beta$ emerge.

As in Sec.\ref{sec:theory},
we define $F(\alpha, \beta) \equiv - \alpha + f(\alpha, \beta)$,
with $f(\alpha, \beta)$ as the right-hand side of Eq.\ref{eq:alpha}.
It's easy to see that $F(\alpha, \beta)|_{\alpha = 0} = 0$.
We further consider the derivative of $F(\alpha, \beta)$
with respect to $\alpha$ at $\alpha = 0$.

\begin{eqnarray}
\frac{\partial F(\alpha, \beta)}{\partial \alpha}
& = &
-1
- p \sum _{k_{+}, k_{-}}
Q_{-}(k_{+}, k_{-})
[\sum _{k_{1} = 0}^{k_{in} - 1}
\left( \begin{array}{c} k_{+} \\ k_{1} \end{array} \right)
[k_{1} \alpha ^{k_{1} - 1} (1 - \alpha)^{k_{+} - k_{1}}
- \alpha ^{k_{1}} (k_{+} - k_{1}) (1 - \alpha)^{k_{+} - k_{1} - 1}]] \nonumber \\
&&
[1 - \sum _{k_{2} = 0}^{k_{ou} - 2} 
\left( \begin{array}{c} k_{-} - 1 \\ k_{2} \end{array} \right)
\beta ^{k_{2}} (1 - \beta) ^{k_{-} - 1 - k_{2}}].
\end{eqnarray}
When $\alpha = 0$,
after some calculation, the second term of the right-hand side of the above equation reduces to $0$,
thus $\frac{\partial F(\alpha, \beta)}{\partial \alpha} |_{\alpha = 0} = -1$.

Summing the above results,
at $\alpha = 0$
we have $F(\alpha, \beta)|_{\alpha = 0} = 0$
and $\frac{\partial F(\alpha, \beta)}{\partial \alpha} |_{\alpha = 0} < 0$.
If a second and also the fixed solution $\alpha ^{*} \neq 0$ for $F(\alpha, \beta) = 0$ emerges
at a critical arc density $c^{*}$ or initial fraction $p^{*}$,
$\alpha ^{*}$ surely shows a gap from $0$.
Correspondingly we have a strictly positive stable $\beta ^{*}$ from Eq.\ref{eq:beta},
and finally a strictly positive $n_{core}^{*}$,
or equivalently a discontinuous emergence of core structure.
An example of the behavior of stable $\alpha$ and $\beta$
on directed ER random graphs
is in figure \ref{fig:kkcore_fb_fa_er}.

\subsubsection{Scaling property}

We further consider the scaling property
above the transition point after the discontinuity happens on directed random graphs.
For the cases with $k_{in} \geq 2$ or $k_{ou} \geq 2$,
the transition is hybrid
with the scaling form $n_{core} - n_{core}^{*} \propto (c - c^{*})^{1/2}$
with the arc density $c$
or $n_{core} - n_{core}^{*} \propto (p - p^{*})^{1/2}$
with the initial fraction $p$.
We first consider the scaling property of the stable $\alpha$
respective to the arc density $c$
above the discontinuous transition point.
The scaling properties of $n_{core}$ respective to $c$ and $p$
follow the same procedure of proof.

For the ease of notation,
we denote the right-hand side of Eq.\ref{eq:alpha} as simply as $f(\alpha)$ instead of $f(\alpha, \beta)$,
along with $F(\alpha)$ instead of $F(\alpha, \beta)$.
At the transition point of the arc density $c^{*}$,
there is an excess degree distribution $Q_{-}^{*}(k_{+}, k_{-})$
which can be denoted as $Q_{-}^{*}$.
Correspondingly,
there is a stable (or degenerate) solution $(\alpha, \beta) = (\alpha ^{*}, \beta ^{*})$
besides the trivial fixed solution $(\alpha, \beta) = (0, 0)$.
At the critical point with $\alpha ^{*}$ and $Q_{-}^{*}$,
we have the relations
$F(\alpha)|_{\alpha = \alpha ^{*}, Q_{-} = Q_{-}^{*}} = 0$ and
$\frac {\partial F(\alpha)}{\partial \alpha} |_{\alpha = \alpha ^{*}, Q_{-} = Q_{-}^{*}} = 0$.
Correspondingly, we have
$f(\alpha ^{*})|_{\alpha = \alpha ^{*}, Q_{-} = Q_{-}^{*}} = \alpha^{*}$ and
$\frac {\partial f(\alpha)}{\partial \alpha} |_{\alpha = \alpha ^{*}, Q_{-} = Q_{-}^{*}} = 1$.
We further expand $f(\alpha)$ slightly above the transition point $\alpha ^{*}$
and the corresponding $Q_{-}^{*}$.

\begin{eqnarray}
\alpha ^{*} + \delta \alpha
& = &
f(\alpha ^{*})|_{\alpha = \alpha ^{*}, Q_{-} = Q_{-}^{*}} \nonumber \\
& + &
\frac {\partial f}{\partial \alpha}|_{\alpha = \alpha^{*}, Q_{-} = Q_{-}^{*}} \delta \alpha
+ \sum _{k_{+}, k_{-}}
\frac {\partial f}{\partial Q_{-}(k_{+}, k_{-})}|_{\alpha = \alpha ^{*}, Q_{-} = Q_{-}^{*}} \delta Q_{-}(k_{+}, k_{-}) \nonumber \\
& + &
\frac {1}{2} \frac {\partial ^{2} f}{\partial \alpha ^{2}}|_{\alpha = \alpha ^{*}, Q_{-} = Q_{-}^{*}}
(\delta \alpha)^2
+ \sum _{k_{+}, k_{-}} \frac {\partial ^{2} f}
{\partial \alpha \partial Q_{-}(k_{+}, k_{-})}|_{\alpha = \alpha^{*}, Q_{-} = Q_{-}^{*}}
\delta \alpha \delta Q_{-}(k_{+}, k_{-})  \nonumber \\
& + &
\frac {1}{2} \sum _{k_{+}, k_{-}, k_{+}^{'}, k_{-}^{'}}
\frac {\partial ^{2} f}{\partial Q_{-}(k_{+}, k_{-}) \partial Q_{-}^{'}(k_{+}^{'}, k_{-}^{'})}
|_{\alpha = \alpha^{*}, Q_{-} = Q_{-}^{*}, Q_{-}^{'} = Q_{-}^{*}}
\delta Q_{-}(k_{+}, k_{-}) \delta Q_{-}^{'}(k_{+}^{'}, k_{-}^{'}) \nonumber \\
& + &
\textrm{higher orders}.
\end{eqnarray}
Ignoring terms with orders higher than the quadratic ones,
we have

\begin{equation}
\sum _{k_{+}, k_{-}} \frac {\partial f}{\partial Q_{-}(k_{+}, k_{-})}
|_{\alpha = \alpha ^{*}, Q_{-} = Q_{-}^{*}}
\delta Q_{-}(k_{+}, k_{-})
+ \frac {1}{2}
\frac {\partial ^{2} f}{\partial \alpha ^{2}}
|_{\alpha = \alpha ^{*}, Q_{-} = Q_{-}^{*}}
(\delta \alpha)^{2}
= 0.
\end{equation}
Then we have

\begin{equation}
\delta \alpha
\approx
|\sum _{k_{+}, k_{-}}
\frac {\partial f}{\partial Q_{-}(k_{+}, k_{-})}
|_{\alpha = \alpha ^{*}, Q_{-} = Q_{-}^{*}}
\delta Q_{-}(k_{+}, k_{-})|^{\frac{1}{2}}.
\end{equation}
For graphs at the critical point $Q_{-}^{*}$
with the corresponding critical arc density $c^{*}$,
we have

\begin{equation}
\delta c
\equiv
c - c^{*}
\approx
\sum _{k_{+}, k_{-}} k_{-} \delta P(k_{+}, k_{-})
\approx c^{*} \sum _{k_{+}, k_{-}} \delta Q_{-}(k_{+}, k_{-}).
\end{equation}
Combining the above two equations, we have $\delta \alpha \propto (\delta c)^{1/2}$
with $\delta c \ll 1$.

In figure \ref{fig:kkcore_scaling},
we show some results above the discontinuous transition points
with the analytical theory
on the directed random networks.

\section{Conclusion}
\label{sec:conclusion}

In this paper,
we analytically study a generalized $k$-core pruning process
as a node failure model
based on both the in-degrees and the out-degrees of nodes
to explore the effect of interaction directions
on the resilience of directed networks.
We test our theory
on uncorrelated directed random networks
as well as on real networks,
and we show analytically
that the introduction of unidirectional interactions between nodes
can drive the networks more prone to abrupt collapses
against a degree-based scheme of failures or damages
which distinguishes in-degrees and out-degrees for nodes.

Here we briefly discuss some related problems worthy of further exploration.
(1) Alternative definitions of node failures and pruning processes.
In the model we consider here,
an intact node has large enough both an in-degree and an out-degree.
Based on this idea,
we define the $(k_{in}, k_{ou})$-core pruning process on directed networks.
Yet other definitions of
node intactness and the corresponding node removal processes
are possible in specific contexts.
For example,
the paper \cite{
AzimiTafreshi.etal-PRE-2013}
considers a generalized way of greedy leaf removal
in which only in-coming or out-going arcs of a node
are involved in a basic removal step;
the paper
\cite{
Radicchi.Binaconi-PRX-2017}
presents a more relaxed definition of node intactness
to explain the adoption of node interdependency in real connected systems.
(2) Dynamical significance of nested $(k_{in}, k_{ou})$-core structures.
On a directed graph instance,
a $(k_{in}, k_{ou})$-core can be derived with the pruning process
from a $(k_{in}^{'}, k_{ou}^{'})$-core
while $k_{in} \geq k_{in}^{'}$ and $k_{ou} \geq k_{ou}^{'}$.
Thus we can derive a procedure to reveal the nested structure of a directed network
with increasing $k_{in}$ and $k_{ou}$.
The dynamical significance of this nested structure,
for example, in the spreading or the information processing,
can be studied on real networks based on some dynamical models.
A similar study line for the $K$-cores in undirected networks can be found in
\cite{
Carmi.etal-PNAS-2007,
Kitsak.etal-NatPhys-2010}.
(3) Network resilience as an optimization problem.
Here we only consider the prediction problem on the sizes
of the residual structure after network failures or damages.
Yet an optimization version of the problem,
removing a minimal number of nodes along with their adjacent arcs or simply only arcs
to disrupt a $(k_{in}, k_{ou})$-core
in a directed network,
is a totally different and probably an NP-hard problem
\cite{
Garey.Johnson-1979}.
Previous studies as
\cite{
Festa.Pardalos.Resende-1999,
Zhou-JStatMech-2016,
Zhao.Zhou-arxiv-2016}
try to optimally remove the strongly connected components, or equivalently $(1, 1)$-cores,
of directed networks,
yet a general statistical-physical framework
leading to an optimal destruction scheme
of any core structure is still lacking.
A possible framework can be
an extension into the case of directed networks
based on a study of dismantling undirected networks
\cite{
Braunstein.etal-PNAS-2016},
in which a static description for a node removal process to reveal $2$-cores
is derived and is further combined with the cavity method
\cite{
Mezard.Montanari-2009}.

\section{Acknowledgements}

The author acknowledges the support by Fondazione CRT under project SIBYL, initiative "La Ricerca dei Talenti."
This work is stimulated from the author's previous collaboration with Hai-Jun Zhou.
Part of the simulation work is performed on HPC Cluster of ITP-CAS.
The author thanks Alfredo Braunstein, Luca Dall'Asta, Hai-Jun Zhou, Anna Muntoni, and Jacopo Bindi
for helpful discussion and reading the manuscript.
The author thanks Alfredo Braunstein for his help in the final format of the manuscript.

\section{Appendix A. The $(k_{in}, k_{ou})$-core percolation on directed random graphs}

We present here the simplified forms of the iterative equations
for the cavity messages $\alpha$ and $\beta$,
the relative sizes $n_{core}$, and the arc densities $c_{core}$
for the $(k_{in}, k_{ou})$-core percolation
on directed Erd\"{o}s-R\'enyi (ER) random graphs,
directed regular random (RR) graphs,
and directed scale-free (SF) networks
generated both with the configurational model
\cite{
Newman.Strogatz.Watts-PRE-2001,
Zhou.Lipowsky-PNAS-2005}
and the static model
\cite{
Goh.Kahng.Kim-PRL-2001,
Catanzaro.PastorSatorras-EPJB-2009}.

The directed ER random graph with an arc density $c$
generated with the method in the main text
has a degree distribution
$P(k_{+}, k_{-}) = P_{+}(k_{+}) P_{-}(k_{-})$,
while $P_{\pm}(k_{\pm}) = e^{-c}c^{k_{\pm}}/k_{\pm}!$.
The excess degree distributions are
$Q_{+}(k_{+}, k_{-}) = Q_{+}(k_{+}) P_{-}(k_{-})$ and
$Q_{-}(k_{+}, k_{-}) = P_{+}(k_{+}) Q_{-}(k_{-})$,
while $Q_{\pm}(k_{\pm}) = e^{-c}c^{k_{\pm} - 1}/(k_{\pm} - 1)!$.
The iterative equations for $\alpha$ and $\beta$, $n_{core}$, and $c_{core}$
can be simplified as below.

\begin{eqnarray}
\alpha
& = & p
[1- e^{- c \alpha} \sum _{k_1 = 0}^{k_{in} - 1} \frac {(c \alpha)^{k_1}}{k_{1}!}]
[1- e^{- c  \beta}  \sum _{k_2 = 0}^{k_{ou} - 2} \frac {(c  \beta)^{k_2}}{k_{2}!}], \\
\beta
& = & p
[1- e^{- c \alpha} \sum _{k_1 = 0}^{k_{in} - 2} \frac {(c \alpha)^{k_1}}{k_{1}!}]
[1- e^{- c \beta}   \sum _{k_2 = 0}^{k_{ou} - 1} \frac {(c  \beta)^{k_2}}{k_{2}!}], \\
n_{core}
& = & p
[1- e^{- c \alpha} \sum _{k_1 = 0}^{k_{in}  - 1} \frac {(c \alpha)^{k_1}}{k_{1}!}]
[1- e^{- c   \beta} \sum _{k_2 = 0}^{k_{ou} - 1} \frac {(c  \beta)^{k_2}}{k_{2}!}], \\
c_{core}
& = &
\frac {c \alpha \beta}{n_{core}}.
\end{eqnarray}
%

A directed RR graph
generated from an undirected RR graph instance with an integer degree $k_{0}$
as in the main text
has an arc density $c (\equiv k_{0} / 2)$ and
a degree distribution
$P(k_{+}, k_{0} - k_{+}) = \left( \begin{array}{c} k_{0} \\ k_{+} \end{array} \right) / 2^{k_{0}}$.
Correspondingly,
its excess degree distributions are
$Q_{+}(k_{+}, k_{0} - k_{+})
= \left( \begin{array}{c} k_{0} - 1 \\ k_{+} - 1 \end{array} \right) / 2^{k_{0} - 1}$ and
$Q_{-}(k_{+}, k_{0} - k_{+})
= \left( \begin{array}{c} k_{0} - 1 \\ k_{0} - k_{+} - 1 \end{array} \right) / 2^{k_{0} - 1}$.
We can simplify the iterative equations for $\alpha$ and $\beta$, and $n_{core}$ as below.

\begin{eqnarray}
\alpha
& = &
p \sum _{k_{+} = 0}^{k_{0}}
\frac {1}{2^{k_{0} - 1}}
\left( \begin{array}{c} k_{0} - 1 \\ k_{0} - k_{+} - 1 \end{array} \right) \nonumber \\
&&
[1 - \sum _{k_{1} = 0}^{k_{in} - 1} 
\left( \begin{array}{c} k_{+} \\ k_{1} \end{array} \right)
\alpha ^{k_{1}} (1 - \alpha) ^{k_{+} - k_{1}}]
[1 - \sum _{k_{2} = 0}^{k_{ou} - 2} 
\left( \begin{array}{c} k_{0} - k_{+} - 1 \\ k_{2} \end{array} \right)
\beta ^{k_{2}} (1 - \beta) ^{k_{0} - k_{+} - 1 - k_{2}}], \\
\beta
& = &
p \sum _{k_{+} = 0}^{k_{0}}
\frac {1}{2^{k_{0} - 1}}
\left( \begin{array}{c} k_{0} - 1 \\ k_{+} - 1 \end{array} \right) \nonumber \\
&&
[1 - \sum _{k_{1} = 0}^{k_{in} - 2} 
\left( \begin{array}{c} k_{+} - 1 \\ k_{1} \end{array} \right)
\alpha ^{k_{1}} (1 - \alpha) ^{k_{+} - 1 - k_{1}}]
[1 - \sum _{k_{2} = 0}^{k_{ou} - 1} 
\left( \begin{array}{c} k_{0} - k_{+} \\ k_{2} \end{array} \right)
\beta ^{k_{2}} (1 - \beta) ^{k_{0} - k_{+} - k_{2}}], \\
n_{core}
& = &
p \sum _{k_{+} = 0}^{k_{0}}
\frac {1}{2^{k_{0}}}
\left( \begin{array}{c} k_{0} \\ k_{+} \end{array} \right) \nonumber \\
&&
[1 - \sum _{k_{1} = 0}^{k_{in} - 1} 
\left( \begin{array}{c} k_{+} \\ k_{1} \end{array} \right)
\alpha ^{k_{1}} (1 - \alpha) ^{k_{+} - k_{1}}]
[1 -\sum _{k_{2} = 0}^{k_{ou} - 1} 
\left( \begin{array}{c} k_{0} - k_{+} \\ k_{2} \end{array} \right)
\beta ^{k_{2}} (1 - \beta) ^{k_{0} - k_{+} - k_{2}}].
\end{eqnarray}
%

A directed SF network instance
generated with the configurational model
with an in-degree exponent $\gamma _{+}$
and an out-degree exponent $\gamma _{-}$
has a degree distribution

\begin{equation}
P(k_{+}, k_{-}) =
\frac { k_{+}^{- \gamma _{+}} }{ \sum _{k_{+}} k_{+}^{- \gamma _{+}} }
\times
\frac { k_{-}^{- \gamma _{-}} }{ \sum _{k_{-}} k_{-}^{- \gamma _{-}} },
\end{equation}
while the summations in the denominators are carried out as
$k_{\pm}^{min} \leq k_{\pm} \leq k_{\pm}^{max}$,
while $k_{+}^{min}$ and $k_{+}^{max}$ are respectively the minimal and maximal in-degrees,
and  $k_{-}^{min}$ and $k_{-}^{max}$ are respectively the minimal and maximal out-degrees
permitted in the graph instance.
This summation rule for $k_{+}$ and $k_{-}$
applies in the following equations for the directed SF networks
generated with the configurational model.
Correspondingly, its mean arc density
$c = \sum _{k_{+}, k_{-}} k_{+} P(k_{+}, k_{-}) = \sum _{k_{+}, k_{-}} k_{-} P(k_{+}, k_{-})$.
The excess degree distributions
can be calculated directly via
$Q_{\pm} (k_{+}, k_{-}) = k_{\pm} P(k_{+}, k_{-}) / c$.
We can further simplify the equations for $\alpha$, $\beta$, and $n_{core}$.

\begin{eqnarray}
\alpha
& = &
p
[1 -
\frac {1}{\sum _{k_{+}} k_{+}^{- \gamma _{+}}}
\sum _{k_{1} = 0}^{k_{in} - 1}
\alpha ^{k_{1}}
\sum _{k_{+}} k_{+}^{- \gamma _{+}}
\left( \begin{array}{c} k_{+} \\ k_{1} \end{array} \right)
(1 - \alpha)^{k_{+} - k_{1}}] \nonumber \\
& &
\frac {1}{c \sum _{k_{-}} k_{-}^{- \gamma _{-}}}
[\sum _{k_{-}} k_{-}^{- \gamma _{-} + 1}
- \sum _{k_{2} = 0}^{k_{ou} - 2} \beta ^{k_{2}}
\sum _{k_{-}}
k_{-}^{- \gamma _{-} + 1}
\left( \begin{array}{c} k_{-} - 1 \\ k_{2} \end{array} \right)
(1 - \beta)^{k_{-}  - 1 - k_{2}}], \\
\beta
& = &
p
\frac {1}{c \sum _{k_{+}} k_{+}^{- \gamma _{+}}}
[\sum _{k_{+}} k_{+}^{- \gamma _{+} + 1}
- \sum _{k_{1} = 0}^{k_{in} - 2}
\alpha ^{k_{1}}
\sum _{k_{+}}
k_{+}^{- \gamma _{+} + 1}
\left( \begin{array}{c} k_{+} - 1 \\ k_{1} \end{array} \right)
(1 - \alpha)^{k_{+}  - 1 - k_{1}}] \nonumber \\
& &
[1 -
\frac {1}{\sum _{k_{-}} k_{-}^{- \gamma _{-}}}
\sum _{k_{2} = 0}^{k_{ou} - 1}
\beta ^{k_{2}}
\sum _{k_{-}} k_{-}^{- \gamma _{-}}
\left( \begin{array}{c} k_{-} \\ k_{2} \end{array} \right)
(1 - \beta)^{k_{-} - k_{2}}], \\
n_{core}
& = &
p
[1 -
\frac {1}{\sum _{k_{+}} k_{+}^{- \gamma _{+}}}
\sum _{k_{1} = 0}^{k_{in} - 1}
\alpha ^{k_{1}}
\sum _{k_{+}} k_{+}^{- \gamma _{+}}
\left( \begin{array}{c} k_{+} \\ k_{1} \end{array} \right)
(1 - \alpha)^{k_{+} - k_{1}}] \nonumber \\
& &
[1 -
\frac {1}{\sum _{k_{-}} k_{-}^{- \gamma _{-}}}
\sum _{k_{2} = 0}^{k_{ou} - 1}
\beta ^{k_{2}}
\sum _{k_{-}} k_{-}^{- \gamma _{-}}
\left( \begin{array}{c} k_{-} \\ k_{2} \end{array} \right)
(1 - \beta)^{k_{-} - k_{2}}].
\end{eqnarray}
%

A directed SF network instance generated by the static model
can be specified by
an arc density $c$,
an in-degree exponent $\gamma _{+}$,
and an out-degree exponent $\gamma _{-}$.
Defining the parameters $\xi _{\pm} \equiv 1 / (\gamma _{\pm} - 1)$,
we have the degree distribution
$P(k_{+}, k_{-}) = P(k_{+}) P(k_{-})$ as

\begin{eqnarray}
P(k_{\pm})
& = &
\frac {1}{\xi _{\pm}}
\frac {(c (1 - \xi _{\pm}) )^{k_{\pm}}}{k_{\pm}!}
\int _{1}^{\infty}
\mathrm{d} t e^{- c (1 - \xi _{\pm}) t} t^{k_{\pm} - 1 - 1 / \xi _{\pm}} \nonumber \\
& = &
\frac {1}{\xi _{\pm}}
\frac {(c (1 - \xi _{\pm}))^{k_{\pm}}}{k_{\pm}!}
E_{- k_{\pm} + 1 + 1/\xi _{\pm}}(c (1 - \xi _{\pm})),
\end{eqnarray}
while the general exponential integral function
$E_{a}(x) \equiv \int _{1}^{\infty} \mathrm{d}t e^{-xt} t^{-\alpha}$.
The excess degree distributions are
$Q_{+} (k_{+}, k_{-}) = Q_{+}(k_{+}) P_{-}(k_{-})$ and
$Q_{-} (k_{+}, k_{-}) = P_{+}(k_{+}) Q_{-}(k_{-})$,
while

\begin{equation}
Q_{\pm} (k_{\pm})
=
(\frac {1}{\xi _{\pm}} - 1)
\frac {(c (1 - \xi _{\pm}))^{k_{\pm} - 1}}{(k_{\pm} - 1)!}
E_{- k_{\pm} + 1 + 1/\xi _{\pm}}(c (1 - \xi _{\pm})).
\end{equation}
We thus have the simplified iterative equations for $\alpha$, $\beta$, and $n_{core}$.

\begin{eqnarray}
\alpha
& = & p
[1 - \sum _{k_{1} = 0}^{k_{in} - 1}
\frac {1}{\xi _{+}}
\frac {(c (1 - \xi _{+}) \alpha )^{k_{1}}}{k_{1}!}
E_{- k_{1} + 1 + \frac {1}{\xi _{+}}}(c (1 - \xi _{+}) \alpha)] \nonumber \\
&&
[1 - \sum _{k_{2} = 0}^{k_{ou} - 2}
(\frac {1}{\xi _{-}} - 1)
\frac {(c (1 - \xi _{-}) \beta)^{k_{2}}}{k_{2}!}
E_{- k_{2} + \frac {1}{\xi _{-}}}(c (1 - \xi _{-}) \beta)], \\
\beta
& = & p
[1 - \sum _{k_{1} = 0}^{k_{in} - 2}
(\frac {1}{\xi _{+}} - 1)
\frac {(c (1 - \xi _{+}) \alpha)^{k_{1}}}{k_{1}!}
E_{- k_{1} + \frac {1}{\xi _{+}}}(c (1 - \xi _{+}) \alpha)] \nonumber \\
&&
[1 - \sum _{k_{2} = 0}^{k_{ou} - 1}
\frac {1}{\xi _{-}}
\frac {(c (1 - \xi _{-}) \beta)^{k_{2}}}{k_{2}!}
E_{- k_{2} + 1 + \frac {1}{\xi _{-}}}(c (1 - \xi _{-}) \beta)], \\
n_{core}
& = & p
[1 - \sum _{k_{1} = 0}^{k_{in} - 1}
\frac {1}{\xi _{+}}
\frac {(c (1 - \xi _{+}) \alpha)^{k_{1}}}{k_{1}!}
E_{- k_{1} + 1 + \frac {1}{\xi _{+}}}(c (1 - \xi _{+}) \alpha)] \nonumber \\
&&
[1 - \sum _{k_{2} = 0}^{k_{ou} - 1}
\frac {1}{\xi _{-}}
\frac {(c (1 - \xi _{-}) \beta)^{k_{2}}}{k_{2}!}
E_{- k_{2} + 1 + \frac {1}{\xi _{-}}}(c (1 - \xi _{-}) \beta)].
\end{eqnarray}

\section{Appendix B. The $K$-core percolation on undirected graphs}

An undirected graph $G = \{V, E\}$
has a node set $V$ ($|V| = N)$ and an undirected edge set $E$ ($|E| = M$),
correspondingly a mean degree $d \equiv 2M / N$.
Before the $K$-core pruning process,
an initial fraction $1- p$ of nodes are randomly chosen and removed along with all their adjacent edges.
We further apply the pruning process on the network
as any node with a degree $< K$ is iteratively removed,
and the $K$-core structure is the residual graph structure.
A detailed analysis of the mean-field theory of the $K$-core percolation on undirected graphs
can be found in papers like
\cite{
Dorogovtsev.Goltsev.Mendes-PRL-2006}.
Yet in order to show an intrinsic similarity between the analytical frameworks
of the $K$-core percolation on undirected networks
and the $(k_{in}, k_{ou})$-core percolation on directed networks,
we present here the main equations
for calculating the relative size and the mean degree of $K$-cores on random graphs,
and also the simplified equations on undirected ER random and RR networks.

For an undirected graph $G$,
the degree distribution $P(k)$
denotes the probability that a randomly chosen node has $k$ adjacent nearest neighbors.
Following a randomly chosen edge to a node,
for example node $i$,
the excess degree distribution $Q(k)$
is the probability that $i$ has $k$ nearest neighbors.
On the graph $G$,
we randomly choose an edge $(i, j)$.
From the node $i$ which is the $K$-core
following the undirected edge $\{i, j\}$ to node $j$,
we define $\alpha$ as the probability that $j$
is also in $K$-core
when $i$ is not considered.
With the assumption of local tree-like structures on sparse graphs
\cite{
Mezard.Montanari-2009},
the self-consistent equation of $\alpha$
can be derived as

\begin{equation}
\label{eq:kcore_alpha}
\alpha
 = 
 p \sum _{k = K}^{\infty} Q(k)
 \sum _{s = K - 1}^{k - 1}
 \left(\begin{array}{c} k - 1 \\ s \end{array}\right)
 \alpha^{s} (1 - \alpha)^{k - 1 - s}.
\end{equation}
The explanation of the self-consistent equation
is quite like the one for $\alpha$
in the $(k_{in}, k_{ou})$-core percolation problem in the main text.
We denote $n_{core}$ as the normalized size of nodes in the $K$-core structure.
With the stable solution of $\alpha$,
we can derive the formula for $n_{core}$ as

\begin{equation}
\label{eq:kcore_ncore}
n_{core}
 =
 p \sum _{k = K}^{\infty} P(k) \sum _{s = K}^{k}
 \left(\begin{array}{c}k \\ s \end{array}\right)
 \alpha^{s} (1 - \alpha)^{k - s}.
\end{equation}
The degree distribution $P_{core}(k)$ for the $K$-core subgraph
can also be calculated with the stable $\alpha$ as

\begin{equation}
P_{core} (k)
= \frac {1}{n_{core}}
p \sum _{s = K}^{\infty} P(s)
 \left(\begin{array}{c} s \\ k \end{array}\right)
\alpha ^{k} (1 - \alpha)^{s - k}.
\end{equation}
The mean degree of the $K$-core can thus be calculated as
\begin{equation}
c_{core}
=
\sum _{k = K}^{\infty}
k P_{core}(k).
\end{equation}
%

On undirected ER random graphs with a mean degree $d$,
we have $P(k) = e^{-d} d^{k}/k!$
and $Q(k) = e^{-d} d^{k - 1}/(k - 1)!$.
We then have the simplified equations for $\alpha$, $n_{core}$, and $c_{core}$ as

\begin{eqnarray}
\alpha
& = &
p [1 - e^{- d \alpha} \sum _{k = 0}^{K - 2} \frac {(d \alpha)^k}{k!}], \\
n_{core}
& = &
p [1 - e^{- d \alpha} \sum _{k = 0}^{K - 1} \frac {(d \alpha)^k}{k!}], \\
c_{core}
& = &
\frac {d \alpha ^2}{n_{core}}.
\end{eqnarray}
%

On undirected RR graphs with an integer degree $k_{0}$,
we have $P(k) = Q(k) = \delta (k - k_{0})$.
We can further have the simplified equations for $\alpha$ and $n_{core}$.

\begin{eqnarray}
\alpha
& = &
p [1 - \sum _{k = 0}^{K - 2}
 \left(\begin{array}{c} k_{0} - 1 \\ k \end{array}\right)
\alpha ^{k} (1 - \alpha)^{k_{0} - 1 - k}],  \\
n_{core}
& = &
p [1 - \sum _{k = 0}^{K - 1}
 \left(\begin{array}{c} k_{0}  \\ k \end{array}\right)
\alpha ^{k} (1 - \alpha)^{k_{0} - k}].
\end{eqnarray}

\end{document}